\documentclass[11pt]{article}

\let\fn\footnote
\renewcommand{\footnote}[1]{\linespread{1.1}\fn{#1}\linespread{1.29}}

\makeatletter\renewcommand{\section}{\@startsection
{section}{1}{\z@}{-3.5ex plus -1ex minus
    -.2ex}{2.3ex plus .2ex}{\bf }}
\makeatletter\renewcommand{\subsection}{\@startsection{subsection}{2}{\z@}{-3.25ex
plus -1ex minus
   -.2ex}{1.5ex plus .2ex}{\it }}
\makeatletter\renewcommand{\subsubsection}{\@startsection{subsubsection}{3}{-2.45ex}{-3.25ex
plus -1ex minus -.2ex}{1.5ex plus .2ex}{\it }}
\renewcommand{\thesection}{\arabic{section}.}
\renewcommand{\thesubsection}{\arabic{section}.\arabic{subsection}.}

\renewcommand{\theequation}{\thesection\arabic{equation}}
\makeatletter \@addtoreset{equation}{section}

\renewenvironment{thebibliography}[1]
     {\baselineskip=16pt plus 2pt minus 1pt
      \section*{\large\refname
        \@mkboth{\MakeUppercase\refname}{\MakeUppercase\refname}}%
     \list{\@biblabel{\@arabic\c@enumiv}}%
           {\settowidth\labelwidth{\@biblabel{#1}}%
            \leftmargin\labelwidth
            \advance\leftmargin\labelsep
            \@openbib@code
            \usecounter{enumiv}%
            \let\p@enumiv\@empty
            \renewcommand\theenumiv{\@arabic\c@enumiv}}%
      \sloppy
      \clubpenalty4000
      \@clubpenalty \clubpenalty
      \widowpenalty4000%
      \sfcode`\.\@m}

\newcommand{\acknowledgements}{\section*{Acknowledgements}
\addcontentsline{toc}{section}{\hspace{0.6cm}{\bf Acknowledgements}}}

\newcommand{\appendices}{\section*{Appendix}\setcounter{subsection}{0}\setcounter{equation}{0}\renewcommand{\thesubsection}{\Alph{subsection}.}
\renewcommand{\theequation}{\thesubsection\arabic{equation}}
\addtocontents{toc}{\vspace{0.2cm}

{\bf Appendices}}
}

\usepackage{epsfig,rotating}
\usepackage{amsmath,amssymb}
\usepackage{amsfonts}
\usepackage{mathrsfs}
\usepackage{bbm}
\usepackage{bm}
\usepackage[vcentermath,enableskew]{youngtab}

\def\slasha#1{\setbox0=\hbox{$#1$}#1\hskip-\wd0\hbox to\wd0{\hss\sl/\/\hss}}

\def\periodb#1{\setbox0=\hbox{$#1$}#1\hskip-\wd0\hbox to\wd0{-}}

\newcommand{\unit}{\mathbbm{1}}   			

\newcommand{\eff}{{\mathrm{eff}}}   			

\newcommand{\CC}{\mathcal{C}}
\newcommand{\CCC}{\mathscr{C}}

\newcommand{\CCD}{\mathscr{D}}

\newcommand{\CF}{\mathcal{F}}

\newcommand{\CH}{\mathcal{H}}

\newcommand{\CL}{\mathcal{L}}

\newcommand{\CO}{\mathcal{O}}

\newcommand{\CZ}{\mathcal{Z}}

\newcommand{\FR}{\mathbbm{R}}     			
\newcommand{\CPP}{{\mathbbm{C}P}}    			
\newcommand{\ah}{\hat{a}}

\newcommand{\Ch}{\hat{C}}

\newcommand{\dd}{\mathrm{d}}     			
\newcommand{\dpar}{\partial}     			
\newcommand{\diag}{{\mathrm{diag}}}     		
\newcommand{\de}{\mathrm{e}}     			
\newcommand{\di}{\mathrm{i}}     			
\newcommand{\eps}{{\varepsilon}}			
\renewcommand{\Re}{\mathrm{Re}}     			
\renewcommand{\Im}{\mathrm{Im}}     			

\newcommand{\bz}{{\bar{z}}}

\newcommand{\eand}{{~~~\mbox{and}~~~}}     		
\newcommand{\ewith}{{~~~\mbox{with}~~~}}

\newcommand{\der}[1]{\frac{\dpar}{\dpar #1}}   		
\newcommand{\tr}{\,\mathrm{tr}\,}     			

\newcommand{\sU}{\mathsf{U}}     			

\newcommand{\sSU}{\mathsf{SU}}

\newcommand{\sSO}{\mathsf{SO}}

\newcommand{\vac}{|0\rangle}
\newcommand{\cav}{\langle 0|}
\newcommand{\remark}[1]{}     				
\def\tyng(#1){\hbox{\tiny$\yng(#1)$}}			
\def\tyoung(#1){\hbox{\tiny$\young(#1)$}}			

\begin{document}
\begin{titlepage}
\begin{flushright}
  DIAS-STP-07-11
\end{flushright}
\vskip 2.0cm
\begin{center}
{\LARGE \bf Fuzzy Scalar Field Theory\\[0.5cm] as a Multitrace Matrix Model}
\vskip 1.5cm
{\Large Denjoe O'Connor and Christian S{\"a}mann}
\setcounter{footnote}{0}
\renewcommand{\thefootnote}{\arabic{thefootnote}}
\vskip 1cm
{\em School of Theoretical Physics\\
Dublin Institute for Advanced Studies\\
10 Burlington Road, Dublin 4, Ireland}\\[5mm]
{Email: {\ttfamily denjoe, csamann@stp.dias.ie}} \vskip
1.1cm
\end{center}
\vskip 1.0cm
\begin{center}
{\bf Abstract}
\end{center}
\begin{quote}
We develop an analytical approach to scalar field theory on the fuzzy sphere based on considering a perturbative expansion of the kinetic term. This expansion allows us to integrate out the angular degrees of freedom in the hermitian matrices encoding the scalar field. The remaining model depends only on the eigenvalues of the matrices and corresponds to a multitrace hermitian matrix model. Such a model can be solved by standard techniques as e.g.\ the saddle-point approximation. We evaluate the perturbative expansion up to second order and present the one-cut solution of the saddle-point approximation in the large $N$ limit. We apply our approach to a model which has been proposed as an appropriate regularization of scalar field theory on the plane within the framework of fuzzy geometry.
\end{quote}\vskip 25mm
{\footnotesize
}

\end{titlepage}

\section{Introduction}

Roughly speaking, the term {\em fuzzy geometry} labels the noncommutative deformation of  Riemannian manifolds which come with a Laplace operator with discrete spectrum. The idea is to truncate this spectrum and subsequently to deform the product on the truncated set of corresponding eigenfunctions in order to arrive at a closed algebra of functions. The most important example of a fuzzy space is certainly the fuzzy sphere, obtained by truncating the set of spherical harmonics $Y_{lm}$ to the subset $l\leq L$ together with the deformation of the product given by
\begin{equation}
 x^i\star x^j-x^j\star x^i\ =\ [x^i\stackrel{\star}{,}x^j]\ \sim\ \di \eps_{ijk} x^k~,
\end{equation}
where $x^i$, $i=1,\ldots ,3$ are coordinates denoting a point on $S^2\subset\FR^3$. This deformation can be traced back, at least, to Berezin \cite{Berezin:1974du}; the geometrical interpretation is essentially due to Madore \cite{Madore:1991bw}. For a comprehensive review of fuzzy geometry, see \cite{Balachandran:2005ew}.

The concrete physical motivation for studying fuzzy geometry is twofold: First, fuzzy spaces arise -- similarly to other noncommutative spaces -- in string theory rather naturally, when certain background fields are turned on \cite{Myers:1999ps}. Second, field theories on fuzzy spaces reduce in general to finite-dimensional matrix models and there are well-defined limits in which the classical field theories on the fuzzy sphere tend to the continuum field theories on the sphere and on both the commutative and the noncommutative plane. Therefore, fuzzy geometry might provide a powerful regularization scheme for quantum field theories, which would possibly have several advantages over a lattice regularization.

A serious obstacle to using fuzzy geometry as a regulator is the fact that the taking of the limits which yield the commutative spaces does not commute with quantization: In \cite{Chu:2001xi}, it was found that a finite ``noncommutative anomaly'' survives the limit from the fuzzy sphere to the commutative sphere. This anomaly in turn yields the well-known UV/IR mixing on the noncommutative plane \cite{Minwalla:1999px} in a certain limit. An approximate analysis of the phase diagram of fuzzy scalar field theory \cite{Steinacker:2005wj} as well as corresponding numerical studies \cite{Martin:2004un} confirm this point. However, modifications of the na{\"i}ve action used for quantization on the fuzzy sphere might reproduce the correct commutative limit \cite{Dolan:2001gn}. 

To improve our understanding of the situation and to study the effects of the proposed modifications, an analytical handle on the model would be desirable. As mentioned above, field theories on fuzzy spaces are matrix models, and for these models, a large set of techniques has been developed over the last decades, which allows one, for example, to evaluate the partition function of certain matrix models exactly. Furthermore, gauge theory on the fuzzy sphere has recently been solved analytically by using localization techniques on a reformulation of the theory \cite{Steinacker:2007iq}. 
Unfortunately, scalar field theory turns out to be too complicated to allow for a complete treatment.

Using the symmetry of the path integral measure, we can however show that the partition function of fuzzy scalar field theory is the same as that of a multitrace hermitian matrix model. Knowing this model explicitly would be very useful, as such a multitrace matrix model can be solved using e.g.\ the saddle-point approximation. As it turns out, a perturbative expansion of the kinetic term allows us to identify the multitraces appearing in the rewritten action order by order in a straightforward (but increasingly tedious) way. It is clear that such an expansion makes sense in a particular region of the parameter space of the theory. Furthermore, this expansion corresponds to the high temperature (or hopping parameter) expansion which has been successfully used in the calculation of the phase diagram of scalar field theory on a two-dimensional lattice; for details, see \cite{Luscher:1987ay} and references therein.

In this paper, we develop the details of our technique and evaluate the perturbative expansion up to second order together with the solution of the thus obtained multitrace matrix model. Explicitly, we find the effective action 
\begin{equation}
 S_\eff\ =\ \xi_2\sum_{i>j}(\lambda_i-\lambda_j)^2+\xi_{(2,2)}\Big(\sum_{i>j}(\lambda_i-\lambda_j)^2\Big)^2+\xi_4\sum_{i>j}(\lambda_i-\lambda_j)^4+\sum_iV(\lambda_i)~,
\end{equation}
where $\lambda_i$ are the eigenvalues of the matrices encoding the real fields on the fuzzy sphere; the exact values of the coefficients $\xi_2$, $\xi_{(2,2)}$ and $\xi_4$ can be read off from equations \eqref{3.8} and \eqref{3.25}. The large $N$ behavior of the coefficients is given in \eqref{A1A2results} and reads as
\begin{equation}
 \xi_2\ \sim\ \frac{N}{2}~,~~~\xi_{2,2}\ =\ 0\eand \xi_{4}\ \sim\ \frac{N^2}{4}~.
\end{equation}
Moreover, we apply the same technique to the proposed modification of fuzzy scalar field theory mentioned above, and compare the results to the ones from the undeformed model.

The outline of this paper is as follows. We start with the construction of the model, comments on various special cases and symmetry considerations in section 2, before we present the actual perturbative expansion in section 3. The solution of the obtained multitrace matrix model in the large $N$ limit is presented in section 4. In section 5, we consider the modified fuzzy scalar field theory and its corresponding multitrace matrix model, and we conclude in section 6. Our Lie algebra conventions, further useful formul\ae{} and some group theoretical background is found in the appendix.

\section{Scalar $\phi^4$-theory on the fuzzy sphere}

\subsection{The fuzzy sphere}

Consider the sphere $S^2\cong \CPP^1$ together with the Laplace operator obtained from the Fubini-Study metric. Its eigenfunctions are the spherical harmonics $Y_{lm}$ with eigenvalues $l(l+1)$ and degeneracy $2l+1$. The quantized version of this space, which is known as the fuzzy sphere, is obtained by truncating the spectrum at a certain value $L$ and deforming the product such that it closes on the corresponding truncated set of eigenfunctions. This yields a finite-dimensional algebra, which approximates the algebra of functions on $S^2$. To be more precise, note that in complex coordinates, all the $Y_{lm}$ with $l\leq L$ can be written as homogeneous polynomials of degree $L$ in both the homogeneous coordinates $z_\alpha$, $\alpha=1,2$, on $\CPP^1$ and their complex conjugates $\bz_\alpha$. As a basis for these polynomials, we choose the monomials
\begin{equation}\label{2.1}
z_{\alpha_1}\ldots z_{\alpha_L}\bz_{\beta_1}\ldots \bz_{\beta_L}~.
\end{equation}
A suitable quantization of this basis is given by\footnote{This particular representation of functions on the fuzzy sphere can be found in \cite{Dolan:2006tx}.}
\begin{equation}\label{opbasis}
\ah^\dagger_{\alpha_1}\ldots \ah_{\alpha_L}^\dagger\vac\cav \ah_{\beta_1}\ldots \ah_{\beta_L}\in \CH_L\otimes \CH_L^\vee~,
\end{equation}
where $\vac$ is the vacuum in the Fock space $\CF$ of two harmonic oscillators with raising and lowering operators satisfying $[\ah_\alpha,\ah^\dagger_\beta]=\delta_{\alpha\beta}$. Evidently, $\CH_L\subset \CF$ denotes the Hilbert space with $L$ excitations and $\CH_L^\vee$ is its dual. The operators \eqref{opbasis} take over the r{\^o}le of the spherical harmonics \eqref{2.1} and form by construction a closed algebra. They are furthermore in a one-to-one correspondence with endomorphisms of sections of the line bundle $\CO(1)^{\otimes L}\cong\CO(L)$ over $\CPP^1$, and thus our quantization prescription is indeed the usual geometric quantization of $\CPP^1$.

Note that the operators \eqref{opbasis} act on the $N=L+1$-dimensional irreducible representation of $\sSU(2)$ and they can thus be translated into $N\times N$ matrices which become hermitian, if we consider operators corresponding to real functions. Thus, scalar field theories on the fuzzy sphere can be written in the form of hermitian matrix models.

Furthermore, it can be shown that the Laplace operator $\Delta$ obtained from the Fubini-Study metric which is equivalent to $\CL_i\CL_i$ with $\CL_i=\di\eps_{ijk}x^j\dpar_k$, where $x^i$ are Cartesian coordinates denoting points on $S^2\subset \FR^3$, turns into the second Casimir $\Ch_2$ in the representation obtained from the quantization procedure. 

Demanding invariance under the isometry group $\sSU(2)$ suggests that after quantization, the integral over the sphere turns into a trace over the integrand. The appropriate normalization is given by the condition that $\int_{S^2} \dd A=4\pi R^2$.
Therefore,
\begin{equation}
 \int_{S^2} \dd A~f \ \rightarrow\  \frac{4\pi R^2}{N}\tr(\hat{f})~.
\end{equation}
In the following, we will omit the hats over operators for convenience.

Altogether, the fuzzy sphere is defined in terms of a matrix algebra approximating the algebra of functions on the commutative sphere and it is characterized by the parameters $N$ and $R$. We will consider various limits of this matrix algebra in section 5.1.

\subsection{The model}

With the prescriptions of the last section, we can directly translate ordinary scalar $\phi^4$-theory on the sphere to the fuzzy picture. Note that as in the commutative case, we scale the Laplacian by a factor of $\frac{1}{R^2}$. Our action is therefore given by \cite{Grosse:1995ar}
\begin{equation}\label{ActionGen}
S\ =\ \gamma\tr\left(\frac{a}{R^2} \Phi C_2\Phi+r\,\Phi^2+g\,\Phi^4\right)\ =\ \gamma\tr\left(-\frac{a}{2R^2}[L_i,\Phi][L_i,\Phi]+r\,\Phi^2+g\,\Phi^4\right)~,
\end{equation}
where we introduced the shorthand notation $\gamma=\frac{4\pi R^2}{N}$. Furthermore, the $L_i$ are the generators of $\sSU(2)$ in the $L+1=N$-dimensional representation of $\sSU(2)$; see appendix C for our Lie algebra conventions.

The partition function of the model reads as 
\begin{equation}\label{PartitionGen}
Z_{a,r,g,N} \ =\ \int \dd \mu_D(\Phi)~ \de^{-\beta S}\ =\ \int \dd \mu_D(\Phi)~ \de^{-\beta\gamma\tr\left(-\frac{a}{2R^2}[L_i,\Phi][L_i,\Phi]+r\,\Phi^2+g\,\Phi^4\right)}~.
\end{equation}
The measure $\dd \mu_D(\Phi)$ denotes the Dyson measure \cite{Dyson:1962es} on the set of hermitian matrices of dimension $N\times N$,
\begin{equation}
\dd \mu_D(\Phi)\ :=\ \prod_{i\leq j}\dd\Re(\Phi_{ij})\prod_{i<j}\dd\Im(\Phi_{ij})~.
\end{equation}
This measure is invariant under the adjoint action of a unitary matrix $\Omega$, $\Phi\rightarrow \Omega\Phi\Omega^\dagger$; the normalization is irrelevant for our purposes. However, we will rely on the fact that under the decomposition
\begin{equation}
 \Phi\ =\ \Omega \Lambda \Omega^\dagger~,~~~\Lambda\ =\ \diag(\lambda_1,\ldots ,\lambda_N)~,
\end{equation}
the Dyson measure splits according to 
\begin{equation}
\int \dd \mu_D(\Phi)\ =\ \int \prod_{i=1}^N \dd \lambda_i~ \Delta^2(\Lambda) \int \dd \mu_H(\Omega)~.
\end{equation}
Here, $\dd \mu_H(\Omega)$ is the Haar measure as defined in appendix A and $\Delta(\Lambda)$ is the Vandermonde determinant
\begin{equation}
 \Delta(\Lambda)\ :=\ \det ([\lambda_i^{j-1}]_{ij})\ =\ \prod_{i>j} (\lambda_i-\lambda_j)~.
\end{equation}

The explicit appearance of the radius $R$ of the sphere is only interesting for approaching the planar commutative and noncommutative limits. We will therefore put $R=1$ in most of our subsequent discussion.

\subsection{The case $a=0$: Hermitian matrix model}

If we put $a=0$ in \eqref{ActionGen}, our model reduces to the well-known hermitian matrix model. This model is exactly solvable \cite{Brezin:1977sv}. Not only can one straightforwardly reduce this model to a model of eigenvalues, thereby reducing $N^2$ integrations to $N$, but one can even give an explicit expression for the partition function in terms of orthogonal polynomials.

Using the decomposition $\Phi=\Omega\Lambda\Omega^\dagger$, the dependence on $\Omega$ drops out and the model turns into
\begin{equation}
\begin{aligned}
Z_{a=0}&\ =\ \int \prod_{i=1}^N \dd \lambda_i~ \Delta^2(\Lambda) \int \dd \mu_H(\Omega)~ \de^{-\beta\gamma\tr\left(r\sum_i \lambda_i^2+g\sum_i \lambda_i^4\right)}\\&\ =\ \int \prod_{i=1}^N \dd \lambda_i ~ \de^{-2 \sum_{i>j}\ln|\lambda_i-\lambda_j|-\beta\gamma\left(r\sum_i \lambda_i^2+g\sum_i \lambda_i^4\right)}~, 
\end{aligned}
\end{equation}
which is the description of a one-dimensional gas of eigenvalues in a quartic potential with a two-body repulsion. To obtain the exact partition function, consider the monic polynomials $p_n$ of degree $n$ normalized according to 
\begin{equation}
\int \dd \lambda~ p_m p_n \de^{-\beta\gamma\tr\left(r\lambda^2+g\lambda^4\right)}\ =\ h_m\delta_{mn}~.
\end{equation}
Then the partition function reads as 
\begin{equation}
 Z\ =\ N!\prod_{i=0}^{N-1}h_i~.
\end{equation}
As it is well-known, the large $N$ limit of this model has a third order phase transition at the curve
\begin{equation}\label{PhaseTransitionMM}
g\ =\ \frac{\gamma}{4N} r^2\ =\ \frac{\pi R^2}{N^2} r^2~.
\end{equation}
for negative values of $r$. At this curve, the double well potential becomes sufficiently deep, so that the eigenvalue distribution splits into two components. This phase transition will be expected to appear also for $a\neq 0$, as argued in \cite{Steinacker:2005wj}.

\subsection{The case $a\neq 0$: The full fuzzy field theory}

With $a\neq 0$, our model \eqref{ActionGen} becomes considerably more difficult than the hermitian matrix model, as the ``external matrices'' $L_i$ do not commute with unitary matrices, and thus prevent us from performing the integration over the angular degrees of freedom encoded in $\Omega$.

In \cite{DiFrancesco:1992cn}, the partition function for the hermitian matrix model with an additional external matrix $A$ and the interaction term $V(A\Phi)$ has been calculated exactly by counting graphs; this solution was reproduced in \cite{Kazakov:1996zm} using the character expansion method. Unfortunately, having three external matrices $L_i$ as in our model \eqref{ActionGen} renders this method essentially useless.

However, one can make the following statements: The kinetic term does not depend on $\tr(\Phi)$ since $C_2 \unit=0$. Therefore, when $\int \dd\mu_H(\Omega) f(\tr(\Phi C_2\Phi))$ is rewritten as a function of the eigenvalues of $\Phi$, it only depends on the differences $(\lambda_i-\lambda_j)$. Since there has to be a permutation symmetry between the different eigenvalues as well as invariance under $\lambda_i\rightarrow -\lambda_i$, the kinetic term is in fact a function of the form
\begin{equation}
 \int \dd\mu_H(\Omega) \de^{-\tr(\Phi C_2\Phi)}\ =\ \exp\left(-\sum_{k,m_1,n_1,\ldots ,m_k,n_k} \xi_{(m_1,n_1)\ldots (m_k,n_k)}\Xi_{2m_1}^{n_1}\ldots\Xi_{2m_k}^{n_k}\right)~,
\end{equation}
where
\begin{equation}
\Xi_{2m}^{n}\ :=\ \big(\Xi_{2m}\big)^{n}\ :=\ \Big(\sum_{i>j}(\lambda_i-\lambda_j)^{2m}\Big)^n~.
\end{equation}
Besides yielding a nice consistency check for our calculations later on, this form offers a physical interpretation: The kinetic term corresponds to an additional attractive\footnote{The force cannot be repulsive, as the scalar field theory on the fuzzy sphere with $g=0$ and $r>0$ is evidently stable for finite $N$; it can, however, be repulsive on short distances.} force between the eigenvalues, which will compete against the repulsive force stemming from the Vandermonde determinant. 

Altogether, we can make the following qualitative predictions: We expect three different phases, each characterized by the dominance of one of the parameters $a,r$ and $g$.

First, there clearly has to be a remnant of the {\em disordered phase} of the matrix model with $a=0$, in which $g$ is dominant. In this phase, $\langle |\tr(\Phi)|\rangle$ is small due to the eigenvalue repulsion. As the eigenvalues are confined in a single well potential around the origin, also $\langle \tr(\Phi^2)\rangle$ is small.

Second, there can be a {\em non-uniform order phase}, in which $r$ strongly influences the shape of the potential and the double well becomes sufficiently deep. Here, $\langle |\tr(\Phi)|\rangle$ is still small, but $\langle \tr(\Phi^2)\rangle$ becomes considerably larger.

Third, there should be a {\em uniform ordered} phase, in which the kinetic term plays a significant r{\^o}le. As the force introduced by this term corresponds to a potential which vanishes for $\Phi\sim \tau_0$, we expect that in the large $N$ limit, the expectation value of $\langle |\tr(\Phi)|\rangle^2\approx \langle \tr(\Phi)^2\rangle$. Surprisingly, such a phase seems to be present even in the absence of the kinetic term \cite{Shimamune:1981qf}.

\subsection{From fuzzy scalar $\phi^4$-theory to multitrace matrix models}

It is easy to see that any hermitian matrix model can be rewritten in terms of multitrace expressions: Since the Dyson measure is invariant under the adjoint action by an arbitrary unitary matrix $\Omega$,
\begin{equation}
\dd \mu_D(\Phi)\ =\ \dd \mu_D(\Omega\Phi \Omega^\dagger)~,
\end{equation}
the part of the action relevant to integration has to be invariant under this action, as well. Formally, we can write:
\begin{equation*}
\int \dd\mu_D(\Phi)~ \de^{-S}\ =\ \int \dd\mu_D(\Phi)~ \de^{-S_\eff}\ewith
\de^{-S_\eff[\Phi]}\ =\ \frac{1}{\mathrm{vol}(\sU(N))}\int \dd \mu_H(\Omega)~ \de^{-S[\Omega \Phi \Omega^\dagger]}~,
\end{equation*}
and $S_\eff$ is clearly invariant under $\Phi\rightarrow \Omega\Phi \Omega^\dagger$. Now the only functions of $\Phi$ invariant under this transformation are products of traces, and we thus have
\begin{equation*}
S_\eff\ =\ \sum_n s_n \tr(\Phi^n)+\sum_{n,m} s_{nm} \tr(\Phi^n)\tr(\Phi^m)+\sum_{n,m,k}s_{nmk}\tr(\Phi^n)\tr(\Phi^m)\tr(\Phi^k)+\ldots~.
\end{equation*}
Such an action is certainly much easier to study than the original one involving an external matrix. Our goal will therefore be to turn our model of fuzzy scalar $\phi^4$-theory at least approximately into such a multitrace matrix model.

Multitrace matrix models became popular at the beginning of the 1990ies, as they were found to arise naturally when higher order curvature terms were included in matrix models describing two-dimensional gravity \cite{Das:1989fq}, see also \cite{Korchemsky:1992tt}. Including these curvature terms amounts to modifying the kinetic term of the matrix model by adding $\tr(A\Phi A\Phi)$, where $A$ is a fixed external matrix. Before this model was solved exactly in \cite{DiFrancesco:1992cn}, the authors of \cite{Das:1989fq} approximated it by the simplest multitrace matrix model of the form
\begin{equation}\label{MTMM}
 S\ =\  a\tr(\Phi^2)\tr(\Phi^2)+r\tr(\Phi^2)+g \tr(\Phi^4)~.
\end{equation}
This matrix model has been considered as a model for string theories with $c>1$ and additional curvature terms like $\int \dd^2\sigma \sqrt{g} R^2$. It was found that the phase diagrams of both theories share many features \cite{Korchemsky:1992tt}.

\subsection{The most general hermitian matrix model}

As a remark, note that one can push the analysis of the structure of hermitian matrix models using the symmetries of the measure even further. The full symmetry group of the Dyson measure $\dd \mu_D(\Phi)$ on the ensemble of $N\times N$ hermitian matrices is in fact $\sSO(N^2)$, as 
\begin{equation}
\dd \mu_D(\Phi) ~ f(\Phi) \ \sim\ \dd^{N^2} \Phi^\mu~ f(\Phi^\mu\tau_\mu)~.
\end{equation}
where $\dd^{N^2}\Phi^\mu$ is the ordinary volume element on $\FR^{N^2}$. This implies that the only relevant part of the action $S$ is the one which is invariant under $\sSO(N^2)$ rotations of the components $\Phi^\mu$, $S_\eff$, and therefore, the action has to be a function of $\tr(\Phi^2)$:
\begin{equation}
\int \dd\mu_D(\Phi)~ \de^{-S}\ =\ \int \dd\mu_D(\Phi)~ \de^{-S_\eff}\ewith
S_\eff\ =\ \sum_n \tilde{s}_n \left(\tr(\Phi^2)\right)^n~.
\end{equation}
In this expansion, however, the large $N$ limit becomes entangled with the rewriting of the model in terms of $\tr(\Phi^2)$, which renders the reformulation useless for most purposes.

\subsection{The toy model: Hermitian $2\times 2$ matrices}

For simplicity, let us consider for a moment the case $L=1$. The matrices $\Phi$ act here in the fundamental representation of $\sSU(2)$, i.e.\ $\Phi$ is a hermitian $2\times 2$ matrix. We expand $\Phi=\Phi^\mu\tau_\mu$ in the generators $\tau_\mu=(\tau_0,\tau_a)$ of $\sSU(2)$; see appendix C for more details on our conventions. Together with the commutation relations $[\tau_a,\tau_b]=\sqrt{2}\di \eps_{abc}\tau_c$, the action \eqref{ActionGen} simplifies to
\begin{equation}
S\ =\ \gamma\left(2a\Phi^a\Phi^a+r \Phi^\mu\Phi^\mu+g\tr((\Phi^\mu\tau_\mu)^4)\right)~.
\end{equation}
Since $\Phi^a\Phi^a=\tr((\Phi-\frac{1}{2}\unit\tr(\Phi))^2)=\tr(\Phi^2)-\frac{1}{2}\tr(\Phi)\tr(\Phi)$, we can rewrite this expression in terms of traces:
\begin{equation}\label{actionL1traces}
S\ =\ \gamma\left((2a+r)\tr(\Phi^2)-a \tr(\Phi)\tr(\Phi)+g \tr(\Phi^4)\right)~.
\end{equation}
As in the case of the pure matrix model, one can decompose $\Phi$ into the product $\Omega \Lambda\Omega^\dagger$, where $\Omega$ is a unitary matrix and $\Lambda=\diag(\lambda_1,\lambda_2)$. The Dyson measure $\dd\mu_D(\Phi)$ will again split into the measure on the space of eigenvalues $\dd \lambda_1\dd \lambda_2$ and the Vandermonde determinant $(\lambda_1-\lambda_2)^2$. In total, the partition function for $L=1$ reads as
\begin{equation}\label{PartFunctL1}
Z_{a,r,g}\ =\ \int \dd\lambda_1\dd\lambda_2~(\lambda_1-\lambda_2)^2~\de^{-\beta\gamma\left(a(\lambda_1-\lambda_2)^2+r(\lambda_1^2+\lambda_2^2)+g(\lambda_1^4+\lambda_2^4)\right)}~.
\end{equation}
We will use this special case for comparison with our perturbative results for general $N$.

\section{Perturbative expansion of the model}

\subsection{Motivation}

As mentioned in the previous section, our goal will be to turn fuzzy scalar field theory into a multitrace matrix model. Since we do not know of any direct route to integrating out the angular degrees of freedom exactly, we will have to resort to perturbative techniques. This approach is also suggested by the lattice analysis of scalar $\phi^4$-theory on $\FR^2$, where a hopping parameter -- or high-temperature -- expansion has been successfully used, see e.g.\ \cite{Luscher:1987ay} and references therein.

\begin{figure}[h]
\center
\begin{picture}(440,150)
\put(20.0,115.0){\makebox(0,0)[c]{$\CCC$}}
\put(25.0,50.0){\makebox(0,0)[c]{$g$}}
\put(195.0,40.0){\makebox(0,0)[c]{$-r$}}
\put(240.0,115.0){\makebox(0,0)[c]{$\CCC$}}
\put(245.0,50.0){\makebox(0,0)[c]{$g$}}
\put(415.0,40.0){\makebox(0,0)[c]{$-r$}}
{\epsfig{file=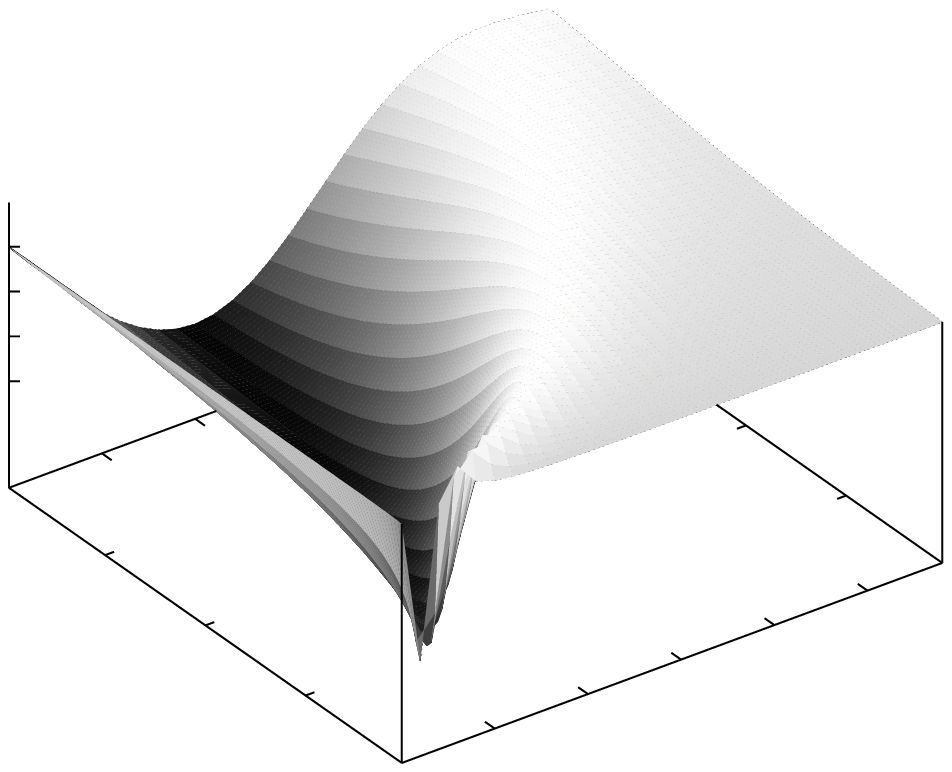, scale=0.6}}
{\epsfig{file=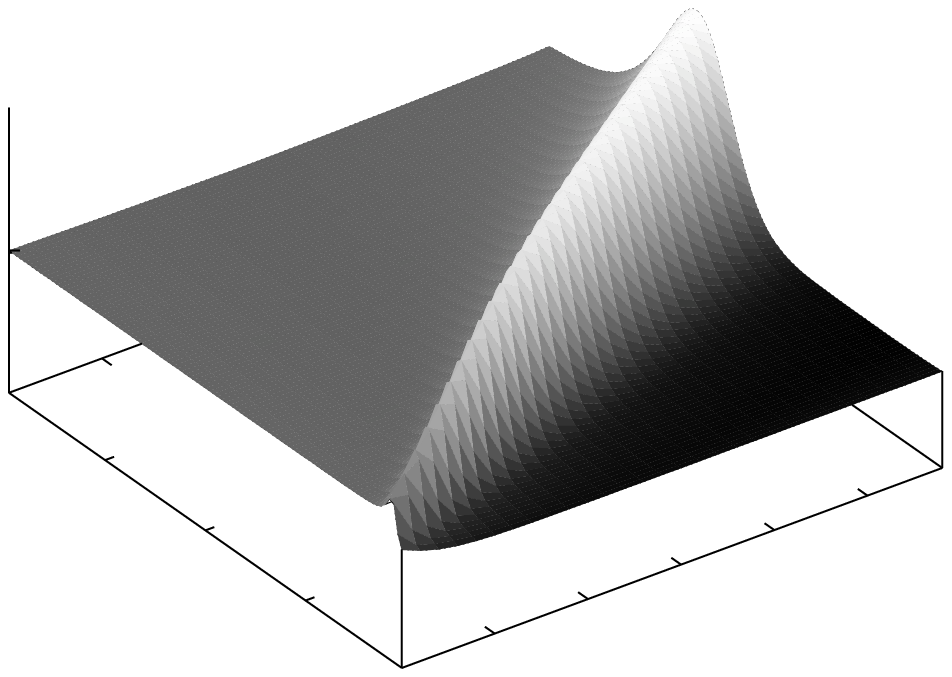, scale=0.6}}
\end{picture}
\caption{On the left, the specific heat $\CCC:=\der{\beta^2}\log Z$ for the pure matrix model with $a=0$ is plotted. The right plot presents the result of the $\CO(a^8)$-approximation to fuzzy $\phi^4$-theory for $N=2$.}\label{fig1}
\end{figure}

Assuming that the coefficient of the kinetic term, $a$, is small -- or equivalently that the coefficients $r$ and $g$ are large, it clearly makes sense to expand the model in powers of $a$. To demonstrate this, we use the high-temperature expansion to compute the specific heat of our model for $N=2$. The results are shown in figure \ref{fig1}, where the specific heat $\CCC$ is plotted in the case $N=2$ for both the pure matrix model and the full theory with high temperature expansion up to $\CO(a^8)$. In the pure matrix model, one clearly sees that the critical line of the third order phase transition is a parabola. The plot corresponding to the full model shows that the critical line for the second order phase transition is indeed a straight line, and the slope extracted from this data deviates from the numerical prediction by only $3.25\%$.

\begin{figure}[h]
\center
\begin{picture}(240,150)
\put(215.0,20.0){\makebox(0,0)[c]{$r$}}
\put(182.0,135.0){\makebox(0,0)[c]{$\CCC$}}
\epsfig{file=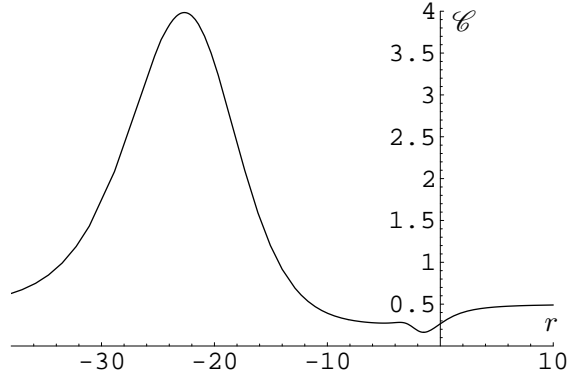}
\end{picture}
\caption{Cross-section through the right plot of figure 1 at $g=50$.}\label{fig2}
\end{figure}

Note that the matrix model phase transition is certainly also present in the full model, however, it is invisible in the second plot, as the scale of the maximum is far larger than the scale of the parabolic valley. In general, for large $g$, the specific heat $\CCC:=\der{\beta^2}\log Z$ interpolates between $\frac{1}{2}$ at large positive and negative $r$. However, there is an intermediate phase, where $\CCC$ approximates $\frac{1}{4}$ which corresponds to the negative $r$ phase of the pure matrix model, see figure \ref{fig2}.

\subsection{Perturbative expansion -- $\CO(a^1)$}

We will use again the Gell-Mann basis and write $\Phi=\Phi^\mu\tau_\mu$. Note that the kinetic term only depends on $\Phi^a$ as $C_2$ annihilates $\tau_0\sim\unit$. We start from the expression
\begin{equation}\label{3.1}
\de^{\beta\gamma a\Phi^aK_{ab}\Phi^b}\ =\ 1+\beta\gamma a \Phi^aK_{ab}\Phi^b+\frac{\beta^2\gamma^2 a^2}{2} \Phi^aK_{ab}\Phi^b\,\Phi^cK_{cd}\Phi^d+\ldots~,
\end{equation}
where
\begin{equation}
K_{ab}\ =\ \tr([L_i,\tau_a][L_i,\tau_b])~.
\end{equation}
As in the case of the pure matrix model, we diagonalize $\Phi=\Omega \Lambda\Omega^\dagger$ using the unitary matrix $\Omega$. This matrix contains the angular information on $\Phi$ and has to be integrated over at the end. We thus have
\begin{equation}
 \Phi^a\ =\ \tr(\tau^a\Omega \Lambda\Omega^\dagger)~,
\end{equation}
and we need to compute 
\begin{equation}\label{3.3}
 \int \dd\mu_H(\Omega)K_{ab}\tr(\tau^a\Omega \Lambda\Omega^\dagger)\tr(\tau^b\Omega \Lambda\Omega^\dagger)
\end{equation}
at first order and
\begin{equation}
 \int \dd\mu_H(\Omega)K_{ab}K_{cd}\tr(\tau^a\Omega \Lambda\Omega^\dagger)\tr(\tau^b\Omega \Lambda\Omega^\dagger)\tr(\tau^c\Omega \Lambda\Omega^\dagger)\tr(\tau^d\Omega \Lambda\Omega^\dagger)
\end{equation}
at second order.

After rewriting these expressions as sums over traces in irreducible representations, we can apply the orthogonality relation\footnote{For a proof of this relation, see appendix A.}
\begin{equation}\label{OrthogonalityBody}
 \int \dd \mu_H(\Omega)~~ [\rho(\Omega)]_{ij}~[\rho^\dagger(\Omega)]_{kl}\ =\ \frac{1}{\dim(\rho)}\delta_{il}\delta_{jk}~,
\end{equation}
to perform the angular integration. We therefore continue by rewriting \eqref{3.3} as 
\begin{equation*}
\tr\big((\tau^a\Omega \Lambda\Omega^\dagger)\otimes(\tau^b\Omega \Lambda\Omega^\dagger)\big)\ =\ \tr\big((\tau^a\otimes\tau^b)(\Omega\otimes\Omega)(\Lambda\otimes \Lambda)(\Omega^\dagger\otimes \Omega^\dagger)\big)\ =:\ \tr([a\otimes b])~.
\end{equation*}
The tensor product\footnote{Note that our notation here obviously deviates from the standard one. For more details on this, see appendix B.} $\tyoung(a)\otimes\tyoung(b)$ splits into $\tyoung(ab)\oplus\tyoung(a,b)$, and thus we have
\begin{equation}
\tr([a\otimes b]) \ =\ \tr_{\tyoung(ab)}([a\otimes b])+\tr_{\tyoung(a,b)}([a\otimes b])~.
\end{equation}
The integral is now easily performed by applying the orthogonality relation \eqref{OrthogonalityBody}, and we arrive at
\begin{equation}
\begin{aligned}
\int\dd\mu_H(\Omega)K_{ab}\Phi^a\Phi^b\ =\ K_{ab}\Big(&\tfrac{1}{\dim(\tyoung(ab))}\tr_{\tyoung(ab)}(\tau^a\otimes \tau^b)\tr_{\tyoung(ab)}(\Lambda\otimes \Lambda)+\\&+\tfrac{1}{\dim(\tyoung(a,b))}\tr_{\tyoung(a,b)}(\tau^a\otimes \tau^b)\tr_{\tyoung(a,b)}(\Lambda\otimes \Lambda)\Big)~.
\end{aligned}
\end{equation}
Note that $\tr_\rho(\Lambda\otimes \Lambda)$ is actually the character of $\Lambda$ in the irreducible representation $\rho$. The traces can be easily computed and the results are given in appendix B. With these formul\ae{}, we obtain
\begin{equation}\label{3.8}
\begin{aligned}
\int\dd\mu_H(\Omega)&K_{ab}\Phi^a\Phi^b\ =\ \tr K\left(\frac{1}{N-N^3}(\tr \Lambda)^2+\frac{1}{N^2-1}\tr \Lambda^2\right)\\
&\ =\ \tfrac{1}{2}N^2\left(\tr \Lambda^2-\frac{1}{N}(\tr \Lambda)^2\right)\ =\ \frac{1}{2}N \sum_{i>j}(\lambda_i-\lambda_j)^2\ =\ \frac{N}{2}\Xi_2~,
\end{aligned}
\end{equation}
where we used
\begin{equation}
\tr(K)\ =\ \tr(C_2)\ =\ \sum_{l=0}^L(2l+1)(l(l+1))\ =\ \tfrac{1}{2}L(1+L)^2(2+L) \ =\ \tfrac{1}{2}(N-1)N^2(N+1)~.
\end{equation}
and
\begin{equation}
\sum_i\lambda_i^2-\tfrac{1}{N}\sum_{i,j}\lambda_i\lambda_j\ =\ \left(1-\tfrac{1}{N}\right)\sum_i\lambda_i^2-\tfrac{2}{N}\sum_{i>j}\lambda_i\lambda_j\ =\ \tfrac{1}{N}\sum_{i>j}(\lambda_i^2-2\lambda_i\lambda_j+\lambda_j^2)~.
\end{equation}

One easily checks that \eqref{3.8} is indeed independent of $\tau_0$: for $\Lambda=\unit_N$ and, equivalently, for eigenvalues $\lambda_1=\ldots =\lambda_N$, the expression vanishes. Note that we can replace $\Lambda$ by $\Phi$ everywhere, keeping the expression independent of the angles $\Omega$. Therefore, one can ``invert'' the integration and replace $K_{ab}\Phi^a\Phi^b$ in \eqref{3.1} under the integral directly according to
\begin{equation}\label{3.10}
\int \dd \mu_D(\Phi)K_{ab}\Phi^a\Phi^b\ =\ \int \dd \mu_D(\Phi)N^2\left(\tr(\Phi^2)-\frac{1}{N}\tr(\Phi)^2\right)~.
\end{equation}

\subsection{Perturbative expansion -- $\CO(a^2)$}

The calculations for higher orders in $a$ and thus $K$ are more and more involved, but nevertheless straightforward. For our approximations, we would like to have one more order in $a$, so let us calculate the involved terms: 
\begin{equation}
 \int \dd \mu_D(\Phi) K_{ab}\Phi^a\Phi^bK_{cd}\Phi^c\,\Phi^d~,
\end{equation}
which turns into 
\begin{equation}
 \int \dd \mu_D(\Phi) \sum_\rho \frac{1}{\dim(\rho)}K_{ab}K_{cd}\tr_\rho(\tau^a\otimes\tau^b\otimes\tau^c\otimes\tau^d)\tr_\rho(\Phi\otimes \Phi\otimes \Phi\otimes \Phi)~.
\end{equation}
after using again the orthogonality relation \eqref{OrthogonalityBody}.

We start by considering the terms $K_{ab}K_{cd}\tr_\rho(\tau^a\otimes\tau^b\otimes\tau^c\otimes\tau^d)$. The traces of this form break up nicely into ordinary traces and multitraces as above, and the detailed formul\ae{} are found in appendix B. Additionally, we use the identities
\begin{equation}
 \begin{aligned}
  K_{ab}K_{cd}\tr(\tau^a\tau^b\tau^c\tau^d)&\ =\ K_{ab}K_{cd}\left(\frac{1}{N}\delta^{ab}\delta^{cd}+\frac{1}{4}d^{abe}d^{cde}\right)~,\\
  K_{ab}K_{cd}\tr(\tau^a\tau^c\tau^b\tau^d)&\ =\ K_{ab}K_{cd}\left(\frac{1}{N}(2\delta^{ac}\delta^{bd}-\delta^{ab}\delta^{cd})+\tfrac{1}{2}d^{ace}d^{bde}-\tfrac{1}{4}d^{bae}d^{ced}\right)~,
 \end{aligned}
\end{equation}
which are easily proven with the identities given in appendix C.

We have now the following contributions in the various representations, multiplicities already taken into account:
\begin{equation}\label{4.16}
\begin{aligned}
 \tyng(4) &\ :\ \frac{1}{24}\left( \left(1+\frac{2}{N}\right) (\tr K)^2+\left(2+\frac{4}{N}\right)\tr K^2+\left(\tfrac{1}{2}d^{abe}d^{cde}+d^{ace}d^{bde}\right)K_{ab}K_{cd}\right)\\
\tyng(3,1) &\ :\ -\frac{1}{8}\left( \left(1+\frac{2}{N}\right) (\tr K)^2+\left(2+\frac{4}{N}\right)\tr K^2+\left(\tfrac{1}{2}d^{abe}d^{cde}+d^{ace}d^{bde}\right)K_{ab}K_{cd}\right)\\ 
\tyng(2,2) &\ :\ \frac{1}{6}\left((\tr K)^2+2\tr K^2\right)\\ 
\tyng(2,1,1) &\ :\ -\frac{1}{8}\left( \left(1-\frac{2}{N}\right) (\tr K)^2+\left(2-\frac{4}{N}\right)\tr K^2-\left(\tfrac{1}{2}d^{abe}d^{cde}+d^{ace}d^{bde}\right)K_{ab}K_{cd}\right)\\ 
\tyng(1,1,1,1) &\ :\ \frac{1}{24}\left( \left(1-\frac{2}{N}\right) (\tr K)^2+\left(2-\frac{4}{N}\right)\tr K^2-\left(\tfrac{1}{2}d^{abe}d^{cde}+d^{ace}d^{bde}\right)K_{ab}K_{cd}\right)
\end{aligned}
\end{equation}
Note that -- as expected -- the sum over the above terms vanishes. To obtain the correction to second order in $a$, we have to divide each term by the dimension of the corresponding representation and multiply by the character of $\Lambda$ (or $\Phi$) in this representation. The latter are taken from table \eqref{TheTable2}, and putting all together, we obtain
\begin{equation}\label{3.16}
\begin{aligned}
\int \dd &\mu_D(\Phi) K_{ab}\Phi^a\Phi^bK_{cd}\Phi^c\,\Phi^d  \ =\ \\ &\int \dd \mu_D(\Phi)
\frac{(2\tr K^2+(\tr K)^2)(t_1^4+8t_1t_3-2(t_1^2t_2+t_4)N+t_2^2(N^2-6)}{N^2(N^4-10N^2+9)}+\\&+\frac{-5t_1^4-4t_1t_3+3t_2^2+(10t_1^2t_2+t_4)N-2(2t_1t_3+t_2^2)N^2+t_4N^3}{N(-36+N^2(-7+N^2)^2)}K\urcorner K
~, 
\end{aligned}
\end{equation}
where we used the shorthand notations
\begin{equation}
 K\urcorner K\ :=\ \left(\tfrac{1}{2}d^{abe}d^{cde}+d^{ace}d^{bde}\right)K_{ab}K_{cd}
\end{equation}
and
\begin{equation}
t_q^p\ :=\ \big(t_q\big)^p\ :=\ \big(\tr(\Lambda^q)\big)^p~.
\end{equation}
As discussed in section 2.4, we can in fact rewrite the integrand in \eqref{3.16} in terms of the two functions:
\begin{equation}
\Xi_4\ =\ \sum_{i>j}(\lambda_i-\lambda_j)^4\eand
\Xi^2_2\ =\ \Big(\sum_{i>j}(\lambda_i-\lambda_j)^2\Big)^2~.
\end{equation}
For this, note that 
\begin{equation}
\begin{aligned}
2\sum_{i>j}(\lambda_i-\lambda_j)^q&\ =\ N t_q-\binom{q}{1}t_{q-1}t_1+\ldots -\binom{q}{q-1}t_{1}t_{q-1}+Nt_q\\&\ =\ \sum_{i=0}^{q}(-1)^{i}(-1)^{q-i}\binom{q}{i}t_{q-i}t_i~.
\end{aligned}
\end{equation}
In our case, $Nt_4$ uniquely identifies the contribution of $\Xi_4$, while $N^2t_2^2$ marks the contribution of $\Xi^2_2$. We thus read off that 
\begin{equation}
\begin{aligned}
\int \dd \mu_D(\Phi) K_{ab}\Phi^a\Phi^bK_{cd}\Phi^c\,\Phi^d  \ =\ &\int \dd \mu_D(\Phi)
\frac{(2\tr K^2+(\tr K)^2)}{N^2(N^4-10N^2+9)}(\alpha_1\Xi_4+\alpha_2\Xi^2_2)+\\&+\frac{1}{N(-36+N^2(-7+N^2)^2)}(\beta_1\Xi_4+\beta_2\Xi^2_2)K\urcorner K
~,
\end{aligned}
\end{equation}
where 
\begin{equation}
\alpha_1\ =\ -2~,~~~\alpha_2\ =\ 1~,~~~\beta_1\ =\ (N-2)(N-3)+5(N-1)~,~~~\beta_2\ =\ -5~;
\end{equation}
performing the complete calculation as a consistency check yields the same result. Using additionally that 
\begin{equation}
\begin{aligned}
(\tr K)^2&\ =\ (\tr C_2)^2\ =\ \tfrac{1}{4}(N-1)^2N^4(N+1)^2~,\\
\tr K^2&\ =\ (\tr C_2^2)\ =\ \tfrac{1}{3}(N-1)^2N^2(N+1)^2~,
\end{aligned}
\end{equation}
we can further simplify our result to
\begin{equation}\label{3.25}
\begin{aligned}
\int \dd \mu_D(\Phi) K_{ab}\Phi^a\Phi^bK_{cd}\Phi^c\,\Phi^d  \ =\ &\int \dd \mu_D(\Phi)
\frac{(N^2-1)(8+3N^2)}{12(N^2-9)}(-2\Xi_4+\Xi^2_2)+\\&+\frac{1}{N(-36+N^2(-7+N^2)^2)}(\beta_1\Xi_4+\beta_2\Xi^2_2)K\urcorner K~.
\end{aligned}
\end{equation}

\subsection{Formula to all orders}

In general, every order in $a$ of the expansion of the kinetic term can be rewritten as a polynomial of the eigenvalues in this manner. Multiplying by the Vandermonde determinant only increases the degree of the polynomial. 

To all orders, the expansion reads as
\begin{equation}
\begin{aligned}
\int\prod_i\dd \lambda_i~ \Delta^2(\Lambda)&\de^{-\beta\gamma\tr(r\,\Phi^2+g\,\Phi^4)} \sum_{n=0}^\infty \frac{(\beta\gamma a)^n}{n!} \sum_{\rho\in Y(2n)} \frac{1}{\dim(\rho)}K_{a_1b_1}\ldots K_{a_nb_n}\times\\&\times\tr_\rho(\tau^{a_1}\otimes\tau^{b_1}\otimes\ldots \otimes\tau^{a_n}\otimes\tau^{b_n})\chi_\rho(\Lambda)~,
\end{aligned}
\end{equation}
where $Y(2n)$ is the set of irreducible representations of $\sSU(N)$ corresponding to Young diagrams with $2n$ boxes, $\Lambda$ is the diagonal matrix of the eigenvalues of $\Phi$ and $\chi_\rho(\Lambda)$ denotes the formula for the character in the representation $\rho$, evaluated at $\Lambda$.

If one were to use orthogonal polynomials for solving the approximated model, note that the only integrals which would appear are of the form
\begin{equation*}
\begin{aligned}
\int \dd \lambda~ \lambda^n\de^{r \lambda^2-g \lambda^4}\ =\ \tfrac{1}{4}(1+(-1)^n)\left(\frac{g}{r^2}\right)^{\frac{3+n}{4}}\left(\sqrt{\frac{g}{r^2}}\Gamma\left(\frac{1+n}{4}\right){}_1F_1\left(\frac{1+n}{4};\frac{1}{2};\frac{g^2}{4r}\right)+\right.\\\left.\Gamma\left(\frac{3+n}{4}\right){}_1F_1\left(\frac{3+n}{4};\frac{3}{2};\frac{g^2}{4r}\right)\right)~,
\end{aligned}
\end{equation*}
and since they are thus exactly soluble for $\frac{g}{r^2}>0$, which corresponds to our case, one can in principle perform exact analytic computations at every order in $a$. In practice, the terms appearing will be more and more involved and it is thus advisable to use a computer algebra program.

\subsection{Verification for $N=2$}

To verify the above description, let us briefly compare to the exact results in the case $N=2$, i.e.\ $L=1$. From \eqref{actionL1traces}, we obtain
\begin{equation}\label{KL1firstorder}
\int \dd \mu_D(\Phi)~\Phi^aK_{ab}\Phi^b\ =\ \int \dd \mu_D(\Phi)~2\tr(\Phi^2)-\tr(\Phi)\tr(\Phi)~,
\end{equation}
which is identical to \eqref{3.10} in the case $N=2$. 

In the case of the term at second order, the general expansion of the tensor product is truncated according to
\begin{equation}
\begin{aligned}
\young(\mu )\otimes\young(\nu )\otimes\young(\rho )\otimes\young(\sigma )~~\ =\ ~~ &\young(\mu \nu\rho\sigma )~\oplus~\young(\mu \nu\rho,\sigma)~\oplus~\young(\mu \nu\sigma,\rho)~\oplus~\young(\mu \rho\sigma ,\nu)~\oplus~\\&
~\oplus~\young(\mu \nu,\rho\sigma )~\oplus~\young(\mu \rho,\nu\sigma)
 \end{aligned} 
\end{equation}
As a consistency check, one can sum $\tr_\rho$ over these representations which yields the usual $\tr(\tau^\mu)\tr(\tau^\nu)\tr(\tau^\rho)\tr(\tau^\sigma)$. Using the identities (valid for $L=1$)
\begin{equation}
\begin{aligned}
K_{ab}K_{cd}\tr(\tau^a\tau^b\tau^c\tau^d)&\ =\ \tfrac{1}{2}\delta^{ab}\delta^{cd}K_{ab}K_{cd}\ =\ \tfrac{1}{2}(\tr K)^2~,\\
K_{ab}K_{cd}\tr(\tau^a\tau^c\tau^b\tau^d)&\ =\ \tfrac{1}{2}(\delta^{ac}\delta^{bd}+\delta^{ad}\delta^{bc}-\delta^{ab}\delta^{cd})K_{ab}K_{cd}\ =\ \tfrac{1}{2}(\tr K)^2~,
\end{aligned}
\end{equation}
and the formul\ae{} given in table \eqref{TheTable}, one easily verifies that 
\begin{equation}
\begin{aligned}
\int \dd \mu_H(\Phi)~\Phi^aK_{ab}\Phi^b\,\Phi^cK_{cd}\Phi^d&\ =\ \int \dd \mu_H(\Phi)~\left(-4 \tr\Phi^4+6(\tr\Phi^2)^2-(\tr\Phi)^4\right)\\
&\ \ =\  \ (\lambda_1-\lambda_2)^4~,
\end{aligned}
\end{equation}
in agreement with the term of order $\CO(a^2)$ in the Taylor expansion of \eqref{PartFunctL1}.

\section{Solution of the model for large $N$}

\subsection{Large $N$ limit}

Numerical results \cite{Martin:2004un} suggest, that the parameters of fuzzy scalar field theories at different $L$ can be rescaled to make the corresponding phase diagrams collapse to one unique diagram. This implies in particular, that we can study the large $N$ limit, in which the formulas simplify and we can apply the saddle-point approximation. Let us therefore extract the large $N$ limit in the following.

Using the identity $\tau^\mu_{ij}\tau^\mu_{kl}=\delta_{il}\delta_{jk}$ and 
\begin{equation}
\begin{aligned}
K_{ab}K_{cd}\tr(\tau^a\tau^b\tau^c\tau^d)&\ =\ \tfrac{1}{4}\tr([L_i,\tau^a][L_i,\tau^b])\tr([L_j,\tau^c][L_j,\tau^d])\tr(\tau^a\tau^b\tau^c\tau^d)\\&\ =\ \tfrac{1}{4}\tr([L_i,\tau^\mu][L_i,\tau^\nu])\tr([L_j,\tau^\rho][L_j,\tau^\sigma])\tr(\tau^\mu\tau^\nu\tau^\rho\tau^\sigma)~,
\end{aligned}
\end{equation}
we find that
\begin{equation}
\begin{aligned}
  K_{ab}K_{cd}\tr(\tau^a\tau^b\tau^c\tau^d)&\ =\ \tfrac{1}{2}N^2\tr(L_iL_iL_jL_j)+\tfrac{1}{2}N(\tr(L_iL_i))^2~,\\
  K_{ab}K_{cd}\tr(\tau^a\tau^c\tau^b\tau^d)&\ =\ \tfrac{1}{2}\tr([L_i,L_j][L_i,L_j])~.
 \end{aligned}
\end{equation}
We have furthermore $L_iL_i=\tfrac{1}{2}(N^2-1)\unit_N$ and $\tr(L_iL_j)=\frac{1}{6}N(N^2-1)\delta_{ij}$, from which it follows that 
\begin{equation}
\begin{aligned}
\tfrac{1}{2}N^2\tr(L_iL_iL_jL_j)+\tfrac{1}{2}N(\tr(L_iL_i))^2&\ =\ \tfrac{1}{4}N^3(N^2-1)^2~,\\
\tfrac{1}{2}\tr([L_i,L_j][L_i,L_j])&\ =\ -N(N^2-1)~.
\end{aligned}
\end{equation}
Thus, the dominant term in \eqref{4.16} is $(\tr K)^2\sim N^8$, in particular the term $K\urcorner K$ is negligible in the large $N$ limit. Altogether, \eqref{3.16} eventually reduces to 
\begin{equation}\label{A1A2results}
\int \dd \mu_D(\Phi) K_{ab}\Phi^a\Phi^bK_{cd}\Phi^c\,\Phi^d  \ =\ \int \dd \mu_D(\Phi) \left(-\frac{N^2}{2} \Xi_4+\frac{N^2}{4} \Xi^2_2\right)~,
\end{equation}
where the second term is precisely the next term in the Taylor expansion of the exponentiated correction in $\CO(a^1)$. The first term is new, and corrects the matrix model. Note that it has the right sign to be exponentiated again and keep the partition function well-defined.

One is clearly tempted to conjecture that at $n$th order in the high-temperature expansion, the term $\sum_{i>j}(\lambda_i-\lambda_j)^{2n}$ is new, while all the other terms just correspond to terms in the Taylor expansion of corrections at smaller orders. This, however, seems not yet possible to prove with the techniques at hand.

If this assumption is correct, we can reexponentiate the terms and arrive at the following matrix model, written in terms of eigenvalues:
\begin{equation}
S\ =\ \gamma\sum_i\left(r\lambda_i^2+g\lambda_i^4\right)+\gamma\sum_{i>j}\left(-\tfrac{a}{2}N(\lambda_i-\lambda_j)^2+\tfrac{\gamma a^2}{4}N^2(\lambda_i-\lambda_j)^4-\tfrac{2}{\gamma}\ln|\lambda_i-\lambda_j|\right)~.
\end{equation}
Note that even if the assumption was not correct, this would still be a useful approximation valid up to order $\CO(a^2)$.

In the large $N$ limit, we rescale $\lambda_i\rightarrow\lambda(\tfrac{i}{N})=\lambda(x)$ with $0<x\leq 1$ and the sums are turned into integrals: $\sum_{i=1}^N\rightarrow N\int_0^1\dd x$. We thus arrive at the action
\begin{equation}
\begin{aligned}
S\ =\ \gamma N \int_0^1\dd x \Big(r\lambda^2(x)+&g\lambda^4(x)+N\int_0^1\dd y \Big(-\tfrac{a}{4}N(\lambda(x)-\lambda(y))^2\\&+\tfrac{\gamma a^2}{8}N^2(\lambda(x)-\lambda(y))^4-\tfrac{1}{\gamma}\ln |\lambda(x)-\lambda(y)|\Big)\Big)~.
\end{aligned}
\end{equation}

\subsection{The saddle-point approximation}

For both single and multitrace hermitian matrix models, the saddle-point approximation provides a powerful tool for analyzing the phase structure. However, one has to integrate out the angular variables first and reduce the matrix model to the eigenvalue formulation. This is due to the fact that the angular degrees of freedom are zero modes in the partition function and keeping them would render the saddle-point approximation invalid.

In the large $N$ limit, we can define scaling limits for the coupling constants, such that all of the terms contribute. From the log-term, we see that we have to pull out a factor of $N^2$. Supporting the validity of our approach is the fact that we can choose a scaling which agrees with the scaling found numerically \cite{Martin:2004un}:
\begin{equation}
a\ =\ N^{\theta_a}\tilde{a}~,~~~r\ =\ N^{\theta_r}\tilde{r}~,~~~g\ =\ N^{\theta_g}\tilde{g}\eand\lambda(x)\ =\ N^{\theta_\lambda}\tilde{\lambda}(x)~,
\end{equation}
where
\begin{equation}
\theta_a\ =\ -\tfrac{1}{2}~,~~~\theta_r\ =\ \tfrac{3}{2}~,~~~\theta_g\ =\ 1\eand \theta_\lambda\ =\ \tfrac{1}{4}~.
\end{equation}
Note that the correct conversion is the following:
\begin{equation}
\theta_{b}\ =\ -1+\theta_r-\theta_a+2\theta_\lambda\eand
\theta_{c}\ =\ -1+\theta_g-2\theta_a+4\theta_\lambda~,
\end{equation}
where $b$ and $c$ are the coupling constant in the normalization of \cite{Martin:2004un}.

Altogether, we have the following partition function:
\begin{equation}
Z\ =\ \int\CCD \lambda~\exp(-N^2\tilde{S})~,
\end{equation}
where 
\begin{equation}\label{effaction}
\begin{aligned}
\tilde{S}\ =\ 4\pi\int_0^1\dd x \Big(\tilde{r}\tilde{\lambda}^2(x)+&\tilde{g}\tilde{\lambda}^4(x)+\int_0^1\dd y \Big(-\tfrac{\tilde{a}}{4}(\tilde{\lambda}(x)-\tilde{\lambda}(y))^2\\&+\tfrac{4\pi\tilde{a}^2}{8}(\tilde{\lambda}(x)-\tilde{\lambda}(y))^4-\tfrac{1}{4\pi}\ln |\tilde{\lambda}(x)-\tilde{\lambda}(y)|\Big)\Big)~.
\end{aligned}
\end{equation}
With our scaling, the saddle-point method becomes exact in the large $N$ limit. 

Rewriting the integral over $y$ in \eqref{effaction} as an integral over $\tilde{\lambda}(y)$ introduces the eigenvalue density $u(\tilde{\lambda})=\dd y/\dd \tilde{\lambda}$, and the moments
\begin{equation}
c_0\ =\ \int \dd \tilde{\mu}~u(\tilde{\mu})\ \stackrel{!}{=}\ 1\eand
c_i\ =\ \int \dd \tilde{\mu}~u(\tilde{\mu}) \tilde{\mu}^i~,~~~i\ =\ 1,\ldots ,4
\end{equation}
appear. Since our problem is symmetric with respect to $\tilde{\lambda}\rightarrow -\tilde{\lambda}$, it is natural to assume that the eigenvalue density will possess this symmetry\footnote{The results of \cite{Shimamune:1981qf} indicate that this is not necessarily so; however, we will postpone the detailed discussion of this point to a future publication.} and therefore the odd moments $c_1$ and $c_3$ have to vanish. Together with suppressing constant terms in $\tilde{\lambda}$, this reduces the action \eqref{effaction} to the following one:
\begin{equation}\label{finaction}
\tilde{S}\ =\ 4\pi\int_0^1\dd x \left(\left(\tilde{r}-\tfrac{\tilde{a}}{4}+\tfrac{6\pi\tilde{a}^2}{2}c_2\right)\tilde{\lambda}^2(x)+\left(\tilde{g}+\tfrac{\pi\tilde{a}^2}{2}\right)\tilde{\lambda}^4(x)-\tfrac{1}{4\pi}\int_0^1\dd y \ln |\tilde{\lambda}(x)-\tilde{\lambda}(y)|\right)~,
\end{equation}
which corresponds to the action of the minimal multitrace matrix model studied in \cite{Das:1989fq} with shifted coefficients. 

\subsection{The solution}

Since the solution to our model can be derived by well-known methods, let us be brief in the calculation. Assuming that we have a one-cut solution, for which $u(\lambda)$ has support on the interval $[-\delta,\delta]$, the saddle-point equation, which is the equation for the stationary points of \eqref{finaction} rewritten using the eigenvalue density $u(\lambda)$ under the constraint $c_0=1$, reads as
\begin{equation}\label{saddlepoint}
 4\pi\left(\tilde{r}-\tfrac{\tilde{a}}{4}+\tfrac{6\pi\tilde{a}^2}{2}c_2\right)\tilde{\lambda}+8\pi\left(\tilde{g}+\tfrac{\pi\tilde{a}^2}{2}\right)\tilde{\lambda}^3\ =\ 
\int_{-\delta}^{\delta} \hspace{-0.619cm}-~~~\dd \tilde{\mu}~\frac{u(\tilde{\mu})}{\tilde{\lambda}-\tilde{\mu}}~,
\end{equation}
where the integral on the right hand side is a principal value integral. The solution to \eqref{saddlepoint}, obtained in the usual manner, is
\begin{equation}\label{evdensity}
u(\tilde{\lambda})\ =\ \left(4\tilde{r}-\tilde{a}+12\pi\tilde{a}^2c_2+4\left(\tilde{g}+\tfrac{\pi\tilde{a}^2}{2}\right)\delta^2+8\left(\tilde{g}+\tfrac{\pi\tilde{a}^2}{2}\right)\tilde{\lambda}^2\right)\sqrt{\delta^2-\tilde{\lambda}^2}~;
\end{equation}
the familiar polynomial deformation of the Wigner semicircle distribution. The normalization condition $c_0=1$ yields
\begin{equation}
\tfrac{1}{2}\pi \delta^2\left(4\tilde{r}+18 \tilde{g}\delta^2-\tilde{a}+3\pi\tilde{a}^2\left(4c_2+3\delta^2\right)\right)\ =\ 1~.
\end{equation}
Solving this equation for $c_2$ results in 
\begin{equation}
c_2\ =\ \frac{2+\pi \delta^2(\tilde{a}-4\tilde{r}-9(2\tilde{g}+\tilde{a}^2\pi )\delta^2)}{12\pi^2\tilde{a}^2\delta^2}~,
\end{equation}
and using the original definition of $c_2$ yields the self-consistency condition
\begin{equation}\label{selfcons}
\pi \delta^2(8\tilde{r}+36\tilde{g}\delta^2+\tilde{a}(3\tilde{a}\pi \delta^2(8+\pi(2\tilde{g}+\tilde{a}^2\pi)\delta^2)-2))-4\ =\ 0~.
\end{equation}

\subsection{The phase diagram}

As discussed before, we expect a non-uniform order phase in the phase diagram, and therefore a region in the parameter space for which the one-cut assumption is no longer valid. In this region, the eigenvalue density \eqref{evdensity} will become negative and therefore ill-defined. Looking at the polynomial part $q(\lambda)$ of $u(\lambda)$, we see that the transition should occur, when $q(0)=0$. Together with the normalization of $c_0$ and the self-consistency condition for $c_2$, this gives a critical curve $\CC$ marking the critical values of $g$ depending on the other parameters. Explicitly, $\CC$ is defined by the following two branches:
\begin{equation}
\CC^\pm\ =\ \frac{\pi}{32}\left(-63\tilde{a}^2-8\tilde{a}\tilde{r}+16\tilde{r}^2\pm(4\tilde{r}-\tilde{a})\sqrt{16 \tilde{r}^2-8\tilde{a}\tilde{r}-95\tilde{a}^2}\right)~.
\end{equation}

The same result can be obtained straightforwardly in a different manner: Let us again assume a symmetric eigenvalue density which allows us to turn to the action
\begin{equation}
\hat{S}\ =\ s_2 \tr(\Phi^2)+s_{2,2}(\tr(\Phi^2))^2+s_4\tr(\Phi^4)~.
\end{equation}
Furthermore, we define the effective coupling $\hat{s}_2=s_2+s_{2,2}\langle\tr(\Phi^2)\rangle$. We can now treat our model as if it was a pure, single-trace hermitian matrix model with $\hat{s}_2$ as the coefficient of the quadratic term. In the regime in which the eigenvalue density has support on two disjoint cuts, we have the usual formula $\langle\tr(\Phi^2)\rangle=-\frac{N \hat{s}_2}{2 s_4}$. Using the equation $\hat{s}_2=-2\sqrt{N s_4}$ for the one-cut two-cut transition, we are lead to the rescaling of fields we used above and eventually arrive at the curve $\CC$ as a self-consistent solution.

To compare this phase transition to the ones obtained numerically, recall that in  \cite{Martin:2004un}, three distinct phases were found.
The transition between the first and the second phase is of third\footnote{A detailed calculation proving this will be presented in a future publication.} order, i.e.\ the specific heat has a discontinuous first derivative at the curve roughly described by equation \eqref{PhaseTransitionMM}. The transition between the second and the third phase is of second order, i.e.\ the specific heat diverges, and numerical results suggest that the critical curve is approximately given by the equation
\begin{equation}\label{critcurv1}
 g\ =\  (0.19\pm0.08)\sqrt{N}r~.
\end{equation}

\begin{figure}[h]
\center
\begin{picture}(440,150)
\put(13.0,130.0){\makebox(0,0)[c]{$g$}}
\put(195.0,20.0){\makebox(0,0)[c]{$-r$}}
\put(230.0,130.0){\makebox(0,0)[c]{$g$}}
\put(415.0,20.0){\makebox(0,0)[c]{$-r$}}
{\epsfig{file=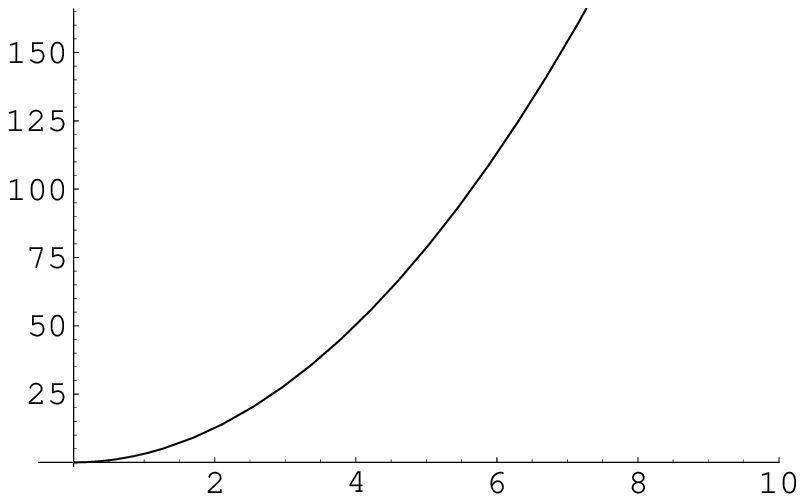, scale=.9}}~~
{\epsfig{file=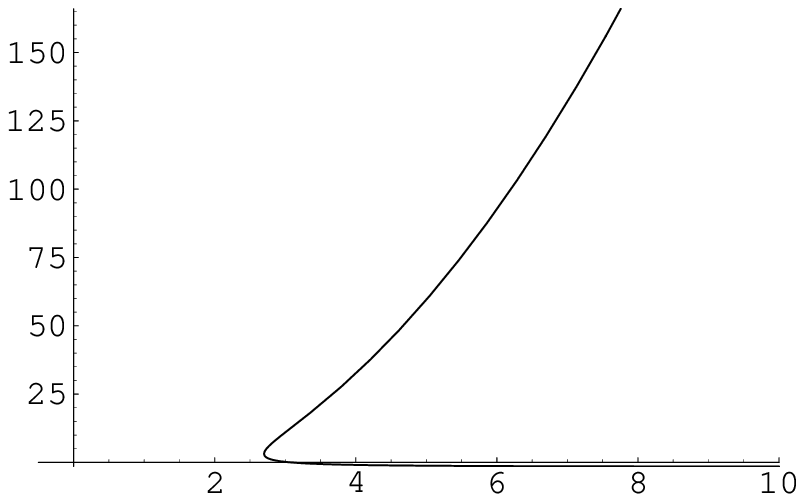, scale=.9}}
\end{picture}
\caption{The one-cut-two-cut transition for the matrix model and the boundary for the region of validity of the one-cut solution to the multitrace matrix model for $\tilde{a}=1$.}\label{fig3}
\end{figure}

The phase boundary we find is a deformation of the matrix model boundary, cf.\ fig. \ref{fig3}. In this deformation, the region in which the one-cut solution is valid is increased by turning on $a$. This is in agreement with the predictions derived from the numerical results, that the triple point at which the third-order phase transition meets the critical line of the second order transition is located at larger values of $r$ than in the case of the pure matrix model transition. Therefore, our results suggest to interpret the turning point of the curve given by $\CC^\pm$ as the rough location of the triple point, and we will use this interpretation in the following section. A comparison with the numerical values confirms this interpretation. The turning point is located at $(r,g)\approx(-2.7,0.25)$ which is to be compared to the result $(r,g)=(-2.3\pm0.2,0.52\pm0.02)$ from the numerical computations \cite{Martin:2004un}, a reasonable agreement for a second order approximation.

We expect the critical curve \eqref{critcurv1} to emerge in a future discussion of the full solution space to our multitrace matrix model.

\section{Modification of the model}

\subsection{$N\rightarrow \infty$ limits of the model}

The various limits with $N\rightarrow \infty$ for scalar field theory on the fuzzy sphere were discussed in \cite{Chu:2001xi}. To make these limits more transparent, note that we can obtain a noncommutative coordinate on $S^2_F$ using the noncommutative analogue of the coordinate $x^i\sim z^\alpha\sigma^i_{\alpha\beta}\bz^\beta$: \begin{equation}
\hat{x}^i\ =\ \ah_\alpha^\dagger\frac{R\sigma^i_{\alpha\beta}}{\sqrt{N^2-1}} \ah_\beta\ewith \left.\hat{x}^i\hat{x}^i\right|_{\CH_{N-1}}\ =\ \left.R^2\hat{\unit}\right|_{\CH_{N-1}}~,
\end{equation}
where $\CH_{N-1}$ is the two-oscillator Hilbert space with $L=N-1$ excitations. These coordinates satisfy the algebra 
\begin{equation}
{}[\hat{x}^i,\hat{x}^j]\ =\ \di \frac{2R}{\sqrt{N^2-1}}\eps_{ijk}\hat{x}^k~.
\end{equation}
There are now the following large $N$ limits:
\begin{itemize}
\setlength{\itemsep}{-1mm}
\item[$\triangleright$] The limit of the commutative sphere $S^2$, for which we simply take $N\rightarrow \infty$ with fixed radius $R$. 
\item[$\triangleright$] The noncommutative planar limit, in which we consider the tangent plane\footnote{A more detailed discussion would use complex coordinates and the stereographic projection.} at the north pole. That is, $\hat{x}^3\rightarrow R$ and therefore the correct limit is $N^2\rightarrow \infty$ and $R^2=\frac{N\theta}{2}$, which results in the noncommutative coordinate algebra $[\hat{x}^1,\hat{x}^2]=\di\theta$ on $\FR^2_\theta$. 
\item[$\triangleright$] The ordinary planar limit, for which $N\rightarrow \infty$ and $R^2\sim N^{1-\eps}$ with $0>\eps>1$. In this limit, the coordinates $\hat{x}^1,\hat{x}^2$ turn into the commutative coordinates on $\FR^2$.
\end{itemize}

\subsection{Commutative and noncommutative $\phi^4$-theory on the plane}

As mentioned in the introduction, the na{\"i}ve action for scalar field theory on the fuzzy sphere will not reproduce standard $\phi^4$-theory on $\FR^2$. There is a UV/IR mixing \cite{Vaidya:2001bt} even in the case of the fuzzy sphere, which is, however, finite for every finite value of $L$ \cite{Chu:2001xi}. In \cite{Dolan:2001gn}, it was shown that the deviation from the commutative theory is given by tadpole diagrams, which can be removed by modifying the action in the following way:
\begin{equation}\label{modify1}
\tilde{S}\ =\ \gamma\tr\left(\frac{a}{R^2} \Phi \left(C_2\CZ_L(C_2)+t-\tfrac{\lambda}{2}R(L,t)\right)\Phi+r\,\Phi^2+g\,\Phi^4\right)~,
\end{equation}
where $\CZ_L(C_2)$ is a power series in $C_2$ corresponding to wave function renormalization, and $R(L,t)$ is a function logarithmically diverging for large $L$.

The numerically obtained structure of the phase diagram of fuzzy scalar field theory \cite{Martin:2004un} suggests a comparison with that of $\phi^4$-theory on $\FR^2$, as the phase transition corresponding to the line is also seen there: The existence of a second order phase transition in scalar $\phi^4$-theory on $\FR^2$ has been rigorously established in \cite{Glimm:1974tz}. This phase transition is due to spontaneously broken reflection symmetry $\phi\rightarrow -\phi$. The exact structure of the phase diagram was calculated only later using lattice and other techniques \cite{Loinaz:1997az}. It was found that for the model
\begin{equation}
\CL_E\ =\ \tfrac{1}{2}(\nabla\phi)^2+\tfrac{1}{2}m^2\phi^2+\tfrac{1}{4}\lambda\phi^4~,
\end{equation}
the phase transition on a lattice occurs along the line $\lambda=-10.24(3)m^2$ in the $\lambda$-$m^2$-plane.

Note that the third order phase transition stemming from the matrix model part together with its triple point with the second order phase transition is absent. This phase finds, however, a nice interpretation in terms of $\phi^4$-theory on the noncommutative plane. There, the theory exhibits a new {\em striped phase}, which breaks translational invariance. This phase has been predicted in \cite{Gubser:2000cd} for $d>2$ using a self-consistent Hartree treatment and in \cite{Chen:2001an} via a one-loop renormalization group analysis. Numerical evidence for the existence of such a new phase has been found in \cite{Ambjorn:2002nj}. As argued in \cite{Bietenholz:2004xs}, appendix B, a definitive confirmation of this result is still pending.

Altogether, we conclude that na{\"i}ve $\phi^4$-theory on the fuzzy sphere serves most likely as a regulator of $\phi^4$-theory on the noncommutative plane. Furthermore, we expect that in the modified model \eqref{modify1}, the triple point moves off towards $r=-\infty$, which would be a clear hint for a fuzzy field theory tending indeed to quantized $\phi^4$-theory on the plane in the appropriate large $N$ and $R$ limits.

\subsection{Modified fuzzy $\phi^4$-theory}

Let us now study the effect of the wave function regularization on the phase diagram, assuming that $\CZ_L(C_2)\approx 1+\kappa C_2$. That is, we consider the modified action
\begin{equation}\label{ActionGenMod}
\tilde{S}\ =\ \gamma\tr\left(\frac{a}{R^2} \Phi (C_2+\kappa C_2C_2)\Phi+r\,\Phi^2+g\,\Phi^4\right)~.
\end{equation}
This modification is particularly simple to realize in our formalism, as it amounts to replacing the kinetic matrix
\begin{equation}
K_{ab} \ \rightarrow\  \check{K}_{ab}\ :=\ K_{ab}+\kappa K_{ac}K_{cb}~.
\end{equation}
We thus have the following new quantities:
\begin{equation}
\begin{aligned}
\tr(\check{K})&\ =\ \tfrac{1}{6}(N-1)N^2(N+1)(3+2(N^2-1)\kappa)~,\\
\tr(\check{K}^2)&\ =\ \tfrac{1}{30}N^2(N^2-1)^2(10+5(3N^2-4)\kappa+2(3N^4-9N^2+8)\kappa^2)~.
\end{aligned}
\end{equation}
To have both contributions in $\check{K}$ survive the large $N$ limit, $\kappa=N^{\theta_\kappa}\tilde{\kappa}$ has obviously to scale with $\theta_\kappa=-2$. After rescaling, we have $\tilde{\check{K}}$, and one can easily show that $\tilde{\check{K}}\urcorner\tilde{\check{K}}$ is still negligible compared to the contribution from $(\tr\tilde{\check{K}})^2$. A straightforward calculation shows that $a$ is simply replaced by 
\begin{equation}
 \check{a}\ =\ a(1+\tfrac{2}{3}\kappa(N^2-1))
\end{equation}
in all equations, and we therefore have in the large $N$ limit
\begin{equation}
 \tilde{\check{a}}\ =\ a(1+\tfrac{2}{3}\tilde{\kappa})~.
\end{equation}
As discussed above, increasing $a$ moves the triple point away from the origin, so we verified that introducing $\kappa$ has the desired effect.

\section{Conclusion}

We gave an explicit algorithm for turning the kinetic term of a fuzzy scalar field theory into a multitrace expression which behaves equivalently under the functional integral. This algorithm uses a perturbative expansion of the kinetic term, and it produces the exact result in a -- in principle -- straightforward manner at every order in the perturbative series. 

Explicitly, we evaluated the multitrace terms for scalar $\phi^4$-theory on the fuzzy sphere up to second order and presented the one-cut solution in the large $N$ limit. We found that this solution gives rise to a consistent deformation of the region of validity of the one-cut assumption in the case of the pure matrix model. Furthermore, this deformation exhibits a turning point, whose location is in good agreement with the location of a triple point conjectured from the numerical data.

Eventually, we studied a modification of the original model which is believed to provide the correct regularization of two-dimensional $\phi^4$-theory within the framework of fuzzy geometry. The multitrace model arising from the deformed theory is a deformation of the multitrace model previously obtained and the observed changes in the phase diagram suggest that the modification is indeed useful.

In a future paper, we will study the full solution space to the multitrace matrix model for both the na{\"i}ve and the modified $\phi^4$-actions. Furthermore, we will try to apply the technique of orthogonal polynomials for solving the model exactly also at finite $N$. Also, we want to probe the multitrace model numerically in order to gain solid data for comparing it with the full fuzzy field theory.

Further directions for future research are also evident: As our rewriting of the kinetic term is independent of the potential, one can easily solve further models with different potentials, as e.g.\ $\phi^6$-theory or even sine-Gordon theory on the fuzzy sphere. Also, field theories on more general fuzzy spaces such as higher-dimensional complex projective spaces \cite{Balachandran:2001dd}, fuzzy flag manifolds \cite{Murray:2006pi} and fuzzy projective algebraic varieties \cite{Saemann:2006gf} are now accessible analytically. In particular, one should study which aspects of non-renormalizability of e.g.\ $\phi^6$-theory in four dimensions remain, when the theory is considered on fuzzy $\CPP^1\times\CPP^1$ or fuzzy $\CPP^2$.

After scalar field theory and Yang-Mills theory on the fuzzy sphere have been approached analytically, it might also be interesting to consider the matrix model corresponding to general relativity on the fuzzy sphere, which has been constructed in \cite{Kurkcuoglu:2006iw}.

Eventually, let us state a rather bold conjecture: Multitrace matrix models were first put forward and studied in the context of trying to define string theories with central charge $c>1$ \cite{Das:1989fq,Korchemsky:1992tt}. It is well-known that such theories exhibit a phase transition with complex critical exponent, an earmark of high instability, and this instability is believed to correspond to a crumpling of the Riemann surfaces. There is a natural class of multitrace matrix models which arises from applying our approach to scalar field theories on the fuzzy sphere and fuzzy Riemann surfaces\footnote{For a definition of such spaces, see e.g.\ \cite{Arnlind:2006ux}. Alternatively, one can define them by considering an embedding of the Riemann surfaces into $\CPP^n$ and using the techniques of \cite{Saemann:2006gf}.} of higher genera. Furthermore, the truncation of the algebra of functions provides a rather natural mechanism for keeping the Riemann surfaces from crumpling in an uncontrolled way. It might indeed be worthwhile to look for a connection between fuzzy field theories and the long-sought definition of $c>1$ string theories.

\acknowledgements
The authors would like to thank Wolfgang Bietenholz, Sergey Cherkis, Francis Dolan, Vladi\-mir Kazakov, Sean Murray, Werner Nahm, Alexander Povolotsky, Harold Steinacker and Sebastian Uhlmann for useful conversations in relation with this work. CS gratefully acknowledges financial support from the Dublin Institute for Advanced Studies. This work was partially supported by MRTN-CT-2006-031962.

\appendices

\subsection{$\sSU(N)$ orthogonality relation}

In the above discussion, we used the orthogonality relation
\begin{equation}\label{OrthogonalityRelation}
 \int \dd \mu_H(\Omega)~~ [\rho(\Omega)]_{ij}~[\rho^\dagger(\Omega)]_{kl}\ =\ \frac{1}{\dim(\rho)}\delta_{il}\delta_{jk}~,
\end{equation}
where $\Omega\in\sSU(N)$, $\rho$ is a finite-dimensional, unitary, irreducible representation and $\rho^\dagger$ denotes its complex conjugate. The measure $\dd \mu_H(\Omega)$ is the Haar measure\footnote{i.e.\ the measure which is invariant under left or right multiplication by $\rho(\Upsilon)$, $\Upsilon\in\sSU(N)$} on $\sSU(N)$ normalized according to $\int \dd \mu_H(\Omega)=1$.

We briefly recall the proof of this relation, as given e.g.\ in \cite{Murnaghan1938}. Consider the integral $I=\int \dd\mu_H(\Omega)\rho(\Omega)A\rho^\dagger(\Omega)$. Using the identity $\rho(\Upsilon)I\rho^\dagger(\Upsilon)=I$, which is a trivial consequence of the Haar measure being invariant under translations $\Omega\rightarrow \Upsilon \Omega$, $\Upsilon\in\sSU(N)$, one can show that $\rho(\Upsilon)I=I\rho(\Upsilon)$. Thus, we conclude by Schur's lemma that $I=c \unit$. Choosing $A_{mn}=\delta_{mj}\delta_{nk}$, we obtain $\int \dd\mu_H(\Omega)[\rho(\Omega)]_{ij}[\rho^\dagger(\Omega)]_{kl}=c_{jk}\delta_{il}$. Tracing this expression over the indices $i,l$ yields that $\dim(\rho) c_{jk}=\delta_{jk}$, from which \eqref{OrthogonalityRelation} follows.

\subsection{Trace formul\ae}

In this appendix, some explicit formul\ae{} for traces of Kronecker products of matrices $\tau^{a}\otimes\tau^{b}\otimes\ldots \otimes \tau^{c}$ projected on certain irreducible representations $\rho$ are given. Such traces will be denoted by $\tr_\rho$. Furthermore, we use the shorthand notation
\begin{equation}
(\mu _1\ldots \mu_m)(\nu _1\ldots \nu_n)\ldots \ :=\ \tr(\tau^{\mu_1}\ldots \tau^{\mu_m})\tr(\tau^{\nu_1}\ldots \tau^{\nu_n})\ldots ~.
\end{equation}
Curly traces $\{\rho\sigma \ldots \}$ denote a sum over all distinct permutations of the enclosed factors. Note that simplifications arise for traceless factors. It should be stressed that our labeled Young diagrams are not the usual ones: The labels in the boxes in the Young diagrams describe the action of the Gell-Mann matrices corresponding to the labels. For example, $\tyoung(\mu \nu)$ means the action of $\tau^\mu\otimes\tau^\nu$ in the representation $\tyng(2)$, which reads as
\begin{equation}
\young(\mu \nu)\,_{ij,kl}\ :=\ 
\tfrac{1}{4}\left(\tau^\mu_{ik}\tau^\nu_{jl}+\tau^\mu_{jk}\tau^\nu_{il}+\tau^\mu_{il}\tau^\nu_{jk}+\tau^\mu_{jl}\tau^\nu_{ik}\right)~.
\end{equation}
The tensor products of the fundamental representations we are interested in decompose in the following way into irreducible representations:
\begin{equation}
 \begin{aligned}
  \young(\mu )\otimes\young(\nu )~~\ =\ ~~ &\young(\mu \nu)~\oplus~\young(\mu ,\nu)\\
\young(\mu )\otimes\young(\nu )\otimes\young(\rho )\otimes\young(\sigma )~~\ =\ ~~ &\young(\mu \nu\rho\sigma )~\oplus~\young(\mu \nu\rho,\sigma)~\oplus~\young(\mu \nu\sigma,\rho)~\oplus~\young(\mu \rho\sigma ,\nu)~\oplus~\\&
~\oplus~\young(\mu \nu,\rho\sigma )~\oplus~\young(\mu \rho,\nu\sigma)~\oplus~\\&~\oplus~\young(\mu \nu,\rho,\sigma)~\oplus~\young(\mu \rho,\nu,\sigma)~\oplus~\young(\mu \sigma,\nu,\rho)~\oplus~\young(\mu ,\nu,\rho,\sigma)
 \end{aligned}
\end{equation}
or, for convenience also written in terms of dimensions:
\begin{equation}
\begin{aligned}
N\cdot N\ =\ & \frac{N^2+N}{2} + \frac{N^2-N}{2}\\
N\cdot N\cdot N\cdot N\ =\ & \frac{N^4+6N^3+11N^2+6N}{24}+3\frac{N^4+2N^3-N^2-2N}{8} +2\frac{N^4-N^2}{12}\\&+3\frac{N^4-2N^3-N^2+2N}{8}+\frac{N^4-6N^3+11N^2-6N}{24}~. 
\end{aligned}
\end{equation}

Altogether, we obtain the following relations between the traces in the irreducible representations appearing in these decompositions and the traces in the fundamental representation:
\begin{equation}\label{TheTable}
\begin{tabular}{|c|c|}
\hline
 $\rho$ & $\tr_\rho$\\
\hline
 \tyoung(\mu \nu) &  $\tfrac{1}{2}\big((\mu )(\nu )+(\mu \nu)\big)$ \\[0.2cm]
 \tyoung(\mu ,\nu) &  $\tfrac{1}{2}\big((\mu )(\nu )-(\mu \nu)\big)$\\[0.2cm]
 \tyoung(\mu \nu\rho\sigma ) &  $\tfrac{1}{4!}\big((\mu )(\nu )(\rho )(\sigma )+\{(\mu )(\nu )(\rho \sigma)\}+\{(\mu )(\nu \rho\sigma )\}+$\\ & $+\{(\mu \nu)(\rho \sigma)\}+\{(\mu \nu\rho\sigma )\}\big)$ \\[0.2cm]
 \tyoung(\mu \nu\rho,\sigma) & $\tfrac{1}{8}\big((\mu )(\nu )(\rho )(\sigma )+\{(\mu \nu)(\rho )\}(\sigma )-(\mu \sigma)(\nu )(\rho )-$\\ & $-(\mu \sigma)(\nu \rho)-(\mu \sigma\{\nu)(\rho \})+\{(\mu \nu\rho)\}(\sigma )-(\mu \sigma\{\nu\rho\})\big)$\\[0.2cm]
 \tyoung(\mu \nu,\rho\sigma ) & $\tfrac{1}{12}\big((\mu )(\nu )(\rho )(\sigma )+(\mu )(\nu )(\rho \sigma)+(\mu \nu)(\rho )(\sigma )-$\\&$-(\mu )(\nu \sigma)(\rho )-(\mu \rho)(\nu )(\sigma )+\{(\mu \nu)(\rho \sigma)\}-(\mu \rho\nu)(\sigma )-(\mu )(\nu \rho\sigma )-$\\ & $-(\mu \sigma\rho)(\nu )-(\mu \nu\sigma)(\rho )-(\mu \sigma\rho\nu)-(\mu \nu\rho\sigma )+(\mu \sigma\nu\rho)+(\mu \rho\nu\sigma)\big)$ \\[0.2cm]
 \tyoung(\mu \nu,\rho,\sigma) & $\tfrac{1}{8}\big((\mu )(\nu )(\rho )(\sigma )-(\nu )\{(\mu )(\rho \sigma)\}+(\mu \nu)(\rho )(\sigma )-(\mu \nu)(\rho \sigma)-$\\ & $-(\mu \nu\{\rho)(\sigma \})+(\mu \{\rho\sigma \})(\nu )+(\mu \nu\{\rho\sigma \})\big)$\\[0.2cm]
 \tyoung(\mu ,\nu,\rho,\sigma) & $\tfrac{1}{4!}\big((\mu )(\nu )(\rho )(\sigma )-\{(\mu )(\nu )(\rho \sigma)\}+\{(\mu )(\nu \rho\sigma )\}+$\\ & $+\{(\mu \nu)(\rho \sigma)\}-\{(\mu \nu\rho\sigma )\}\big)$ \\[0.2cm]
\hline
\end{tabular}
\end{equation}
Note that adding $\tr_\rho$ for all the irreducible representations $\rho$ with two and four boxes yields
\begin{equation}
\begin{aligned}
\tr(\tau^\mu)\tr(\tau^\nu)&\ =\ \tr_{\tyng(1)\otimes \tyng(1)}(\tau^\mu\otimes\tau^\nu)~,\\
\tr(\tau^\mu)\tr(\tau^\nu)\tr(\tau^\rho)\tr(\tau^\sigma)&\ =\ \tr_{\tyng(1)\otimes \tyng(1)\otimes \tyng(1)\otimes \tyng(1)}(\tau^\mu\otimes\tau^\nu\otimes\tau^\rho\otimes\tau^\sigma)~,
\end{aligned}
\end{equation}
respectively, which confirms that our decomposition is correct.

In the case in which all the factors of the Kronecker product are the same, the formul\ae{} given in \eqref{TheTable} reduce to character formul\ae{}, which are found e.g.\ in \cite{Bars:1980yy}. For convenience, we also list them in the following table:
\begin{equation}\label{TheTable2}
\begin{tabular}{|c|c|}
\hline
 $\rho$ & $\chi_\rho(\Lambda)$\\
\hline
 \tyng(2) & $\tfrac{1}{2}\big((\tr \Lambda)^2+\tr \Lambda^2\big)$ \\[0.2cm]
 \tyng(1,1) & $\tfrac{1}{2}\big((\tr \Lambda)^2-\tr \Lambda^2\big)$\\[0.2cm]
\tyng(4) &  $\tfrac{1}{4!}\big((\tr(\Lambda))^4+6(\tr \Lambda)^2\tr \Lambda^2+8\tr \Lambda\tr \Lambda^3+3 (\tr \Lambda^2)^2+6 \tr \Lambda^4\big)$\\[0.2cm]
 \tyng(3,1) & $\tfrac{1}{8}\big((\tr(\Lambda))^4+2(\tr \Lambda)^2\tr \Lambda^2- (\tr \Lambda^2)^2-2 \tr \Lambda^4\big)$\\[0.2cm]
 \tyng(2,2) &$\tfrac{1}{12}\big((\tr(\Lambda))^4-4\tr \Lambda\tr \Lambda^3+3 (\tr \Lambda^2)^2\big)$\\[0.3cm]
 \tyng(2,1,1) & $\tfrac{1}{8}\big((\tr(\Lambda))^4-2(\tr \Lambda)^2\tr \Lambda^2- (\tr \Lambda^2)^2+2 \tr \Lambda^4\big)$\\[0.3cm]
 \tyng(1,1,1,1) &  $\tfrac{1}{4!}\big((\tr(\Lambda))^4-6(\tr \Lambda)^2\tr \Lambda^2+8\tr \Lambda\tr \Lambda^3+3 (\tr \Lambda^2)^2-6 \tr \Lambda^4\big)$\\[0.5cm]
\hline
\end{tabular}
\end{equation}

\subsection{Lie algebra conventions}

We use generators $\tau^\mu=(\tau^0,\tau^a)$ of $\sU(N)$, which are orthonormal, i.e.\
\begin{equation}
 \tr(\tau^\mu\tau^\nu)\ =\ \delta^{\mu\nu}~;
\end{equation}
$\tau^0=\tfrac{1}{N}\unit_N$, and $\tau^a$ are the Gell-Mann matrices generating $\sSU(N)$. In general, we use the Greek letters $\mu,\nu,\rho,\sigma=0,\ldots ,N^2-1$ and Latin letters $a,b,c,d=1,\ldots ,N^2-1$ for labeling generators. The structure constants $f^{abc}$ and the symmetric tensor $d^{abc}$ are defined by the multiplication rule
\begin{equation}
\tau^a\tau^b\ =\ \frac{1}{\sqrt{N}}\delta^{ab}\tau^0+\frac{1}{2}(\di f^{abc}+d^{abc})\tau^c~.
\end{equation}
This implies that 
\begin{equation}
\begin{aligned}
f^{abc}&\ =\ -\di\tr([\tau^a,\tau^b]\tau^c)~~\Rightarrow~~ [\tau^a,\tau^b]\ =\ \di f^{abc}\tau^c~,\\
d^{abc}&\ =\ \tr(\{\tau^a,\tau^b\}\tau^c)~~\Rightarrow~~
\{\tau^a,\tau^b\}\ =\ \frac{2}{\sqrt{N}} \delta^{ab}\tau^0+d^{abc}\tau^c~.
\end{aligned}
\end{equation}
From the above definitions, it is obvious that $f^{abc}$ and $d^{abc}$ are totally antisymmetric and symmetric in their indices, respectively. Recall also the Fierz identity $\tau^\mu_{ij}\tau^\mu_{kl}=\delta_{il}\delta_{jk}$ implying that $\tau^\mu\tau^\mu=N\unit_N$.

The quadratic Casimir $C_2$ with the standard eigenvalues $l(l+1)$ corresponds in our conventions to
\begin{equation}
C_2 \Phi \ =\ \tfrac{1}{2}[\tau_a,[\tau_a,\Phi]]~.
\end{equation}

Starting from
\begin{equation}
 [\tau^\rho,[\tau^\mu,\tau^\nu]]\ =\ \{\tau^\nu\{\tau^\rho,\tau^\mu\}\}-\{\tau^\mu\{\tau^\nu,\tau^\rho\}\}~,
\end{equation}
one easily proves the identity
\begin{equation}\label{Rel_ff}
 f_{ace}f_{bde}\ =\ \frac{4}{N}(\delta_{ba}\delta_{cd}-\delta_{bc}\delta_{ad})+(d_{bae}d_{ced}-d_{bce}d_{aed})~.
\end{equation}
Furthermore, we have
\begin{equation}\label{Rel_fd}
 f_{ace}d_{bde}\ =\ \tr(\tau^a\tau^c\tau^b\tau^d)-\tr(\tau^c\tau^a\tau^b\tau^d)+\tr(\tau^a\tau^c\tau^d\tau^b)-\tr(\tau^c\tau^a\tau^d\tau^b)~.
\end{equation}



\begin{thebibliography}{10}

\bibitem{Berezin:1974du}
F.~A.~Berezin,
{\em General concept of quantization,}
Commun. Math. Phys. {\bf 40} (1975)  153.

\bibitem{Madore:1991bw}
J.~Madore,
{\em The fuzzy sphere,}
Class. Quant. Grav.  {\bf 9} (1992) 69.

\bibitem{Balachandran:2005ew}
A.~P.~Balachandran, S.~Kurkcuoglu, and S.~Vaidya,
{\em Lectures on fuzzy and fuzzy SUSY physics,}
hep-th/0511114.

\bibitem{Myers:1999ps}
R.~C.~Myers,
{\em Dielectric-branes,}
JHEP {\bf 12} (1999)  022 [hep-th/9910053].

\bibitem{Chu:2001xi}
C.-S.~Chu, J.~Madore, and H.~Steinacker,
{\em Scaling limits of the fuzzy sphere at one loop,}
JHEP {\bf 08} (2001)  038 [hep-th/0106205].

\bibitem{Minwalla:1999px}
S.~Minwalla, M.~Van~Raamsdonk, and N.~Seiberg,
{\em Noncommutative perturbative dynamics,}
JHEP {\bf 02} (2000)  020 [hep-th/9912072].

\bibitem{Steinacker:2005wj}
H.~Steinacker,
{\em A non-perturbative approach to non-commutative scalar field theory,}
JHEP {\bf 03} (2005)  075 [hep-th/0501174];
{\em Quantization and eigenvalue distribution of noncommutative scalar field
  theory,}
hep-th/0511076.

\bibitem{Martin:2004un}
X.~Martin,
{\em A matrix phase for the $\phi^4$ scalar field on the fuzzy sphere,}
JHEP {\bf 04} (2004)  077 [hep-th/0402230];
F.~Garcia~Flores, D.~O'Connor, and X.~Martin,
{\em Simulating the scalar field on the fuzzy sphere,}
PoS LAT {\bf 2005} (2005)  262 [hep-lat/0601012];
M.~Panero,
{\em Numerical simulations of a non-commutative theory: the scalar model on the
  fuzzy sphere,}
JHEP {\bf 0705} (2007) 082
[hep-th/0608202].

\bibitem{Dolan:2001gn}
B.~P.~Dolan, D.~O'Connor, and P.~Presnajder,
{\em Matrix $\phi^4$ models on the fuzzy sphere and their continuum limits,}
JHEP {\bf 03} (2002)  013 [hep-th/0109084];
{\em Matrix models on the fuzzy sphere,}
hep-th/0204219.

\bibitem{Steinacker:2007iq}
H.~Steinacker and R.~J.~Szabo,
{\em Localization for Yang-Mills theory on the fuzzy sphere,}
hep-th/0701041.

\bibitem{Luscher:1987ay}
M.~L{\"u}scher and P.~Weisz,
{\em Scaling laws and trivality bounds in the lattice $\phi^4$ theory. 1. One
  component model in the symmetric phase,}
Nucl. Phys. B {\bf 290} (1987) ~25.

\bibitem{Dolan:2006tx}
B.~P.~Dolan, I.~Huet, S.~Murray, and D.~O'Connor,
{\em Noncommutative vector bundles over fuzzy $CP^N$ and their covariant derivatives,}
JHEP {\bf 0707} (2007) 007
[hep-th/0611209].

\bibitem{Grosse:1995ar}
H.~Grosse, C.~Klimcik, and P.~Presnajder,
{\em Towards finite quantum field theory in noncommutative geometry,}
Int. J. Theor. Phys. {\bf 35} (1996)  231 [hep-th/9505175].

\bibitem{Dyson:1962es}
F.~J.~Dyson,
{\em Statistical theory of the energy levels of complex systems. I,}
J. Math. Phys. {\bf 3} (1962)  140.

\bibitem{Brezin:1977sv}
E.~Brezin, C.~Itzykson, G.~Parisi, and J.~B.~Zuber,
{\em Planar diagrams,}
Commun. Math. Phys. {\bf 59} (1978) ~35.

\bibitem{DiFrancesco:1992cn}
P.~Di~Francesco and C.~Itzykson,
{\em A generating function for fatgraphs,}
Ann. Poincare {\bf 59} (1993)  117 [hep-th/9212108].

\bibitem{Kazakov:1996zm}
V.~A.~Kazakov, M.~Staudacher, and T.~Wynter,
{\em Exact solution of discrete two-dimensional $R^2$ gravity,}
Nucl. Phys. B {\bf 471} (1996)  309 [hep-th/9601069].

\bibitem{Shimamune:1981qf}
Y.~Shimamune,
{\em On the phase structure of large $N$ matrix models,}
Phys. Lett. B {\bf 108} (1982)  407.

\bibitem{Das:1989fq}
S.~R.~Das, A.~Dhar, A.~M.~Sengupta, and S.~R.~Wadia,
{\em New critical behavior in $d=0$ large $N$ matrix models,}
Mod. Phys. Lett. A {\bf 5} (1990)  1041.

\bibitem{Korchemsky:1992tt}
G.~P.~Korchemsky,
{\em Matrix model perturbed by higher order curvature terms,}
Mod. Phys. Lett. A {\bf 7} (1992)  3081 [hep-th/9205014];
G.~P.~Korchemsky,
{\em Loops in the curvature matrix model,}
Phys. Lett. B {\bf 296} (1992)  323 [hep-th/9206088];
S.~S.~Gubser and I.~R.~Klebanov,
{\em A modified $c=1$ matrix model with new critical behavior,}
Phys. Lett. B {\bf 340} (1994) ~35 [hep-th/9407014].

\bibitem{Vaidya:2001bt}
S.~Vaidya,
{\em Perturbative dynamics on fuzzy $S^2$ and $RP^2$,}
Phys. Lett. B {\bf 512} (2001)  403 [hep-th/0102212].

\bibitem{Glimm:1974tz}
J.~Glimm and A.~M.~Jaffe,
{\em The $\phi^4_2$ quantum field model in the single-phase region:
  Differentiability of the mass and bounds on critical exponents,}
Phys. Rev. D {\bf 10} (1974)  536;
J.~Glimm, A.~M.~Jaffe, and T.~Spencer,
{\em Phase transition for $\phi^4_2$ quantum fields,}
Commun. Math. Phys. {\bf 45} (1975)  203.

\bibitem{Loinaz:1997az}
W.~Loinaz and R.~S.~Willey,
{\em Monte Carlo simulation calculation of critical coupling constant for
  continuum $\phi^4_2$,}
Phys. Rev. D {\bf 58} (1998)  076003 [hep-lat/9712008];
D.~Lee,
{\em Introduction to spherical field theory,}
Phys. Lett. B {\bf 439} (1998) ~85 [hep-th/9811117].

\bibitem{Gubser:2000cd}
S.~S.~Gubser and S.~L.~Sondhi,
{\em Phase structure of non-commutative scalar field theories,}
Nucl. Phys. B {\bf 605} (2001)  395 [hep-th/0006119].

\bibitem{Chen:2001an}
G.-H.~Chen and Y.-S.~Wu,
{\em Renormalization group equations and the Lifshitz point in noncommutative
  Landau-Ginsburg theory,}
Nucl. Phys. B {\bf 622} (2002)  189 [hep-th/0110134].

\bibitem{Ambjorn:2002nj}
J.~Ambjorn and S.~Catterall,
{\em Stripes from (noncommutative) stars,}
Phys. Lett. B {\bf 549} (2002)  253 [hep-lat/0209106].

\bibitem{Bietenholz:2004xs}
W.~Bietenholz, F.~Hofheinz and J.~Nishimura,
{\em Phase diagram and dispersion relation of the non-commutative $\lambda \phi^4$ model in $d=3$,}
JHEP {\bf 0406} (2004) 042 
[hep-th/0404020].

\bibitem{Balachandran:2001dd}
A.~P.~Balachandran, B.~P.~Dolan, J.-H.~Lee, X.~Martin, and D.~O'Connor,
{\em Fuzzy complex projective spaces and their star-products,}
J. Geom. Phys. {\bf 43} (2002)  184 [hep-th/0107099].

\bibitem{Murray:2006pi}
S.~Murray and C.~Saemann,
{\em Quantization of flag manifolds and their supersymmetric extensions,}
hep-th/0611328.

\bibitem{Saemann:2006gf}
C.~Saemann,
{\em Fuzzy toric geometries,}
hep-th/0612173.

\bibitem{Kurkcuoglu:2006iw}
S.~Kurkcuoglu and C.~Saemann,
{\em Drinfeld twist and general relativity with fuzzy spaces,}
Class. Quant. Grav. {\bf 24} (2007)  291 [hep-th/0606197].

\bibitem{Arnlind:2006ux}
J.~Arnlind, M.~Bordemann, L.~Hofer, J.~Hoppe, and H.~Shimada,
{\em Fuzzy Riemann surfaces,}
hep-th/0602290.

\bibitem{Murnaghan1938}
F.~D.~Murnaghan,
{\em The theory of group representations,}
The Johns Hopkins Press, Baltimore (1938).

\bibitem{Bars:1980yy}
I.~Bars,
{\em U(N) integral for the generating functional in lattice gauge theory,}
J. Math. Phys. {\bf 21} (1980)  2678.

\end{thebibliography}

\end{document}